\documentclass[aps,prb,onecolumn,groupedaddress,10pt]{revtex4}
\usepackage{graphicx,bm,wasysym,color}

\def\be{\begin{equation}} 
\def\ee{\end{equation}}  
\def\ba{\begin{eqnarray}}  
\def\ea{\end{eqnarray}}  
\def\bc{\begin{center}}  
\def\ec{\end{center}}  
\def\p{\partial}

\def\rot{{\rm rot}}

\begin{document}
\title{Third harmonic generation from graphene lying on different substrates: Optical-phonon resonances and interference effects}

\author{N. A. Savostianova} 
\affiliation{Institute of Physics, University of Augsburg, D-86135 Augsburg, Germany}
\author{S. A. Mikhailov}
\email[Email: ]{sergey.mikhailov@physik.uni-augsburg.de}
\affiliation{Institute of Physics, University of Augsburg, D-86135 Augsburg, Germany}

\date{\today}

\begin{abstract}
Graphene is a nonlinear material which can be used as a saturable absorber, frequency mixer and frequency multiplier. We investigate the third harmonic generation from graphene lying on different substrates, consisting of a dielectric (dispersionless or polar), metalized or non-metalized on the back side. We show that the third harmonic intensity emitted from graphene lying on a substrate, can be increased by orders of magnitude as compared to the isolated graphene, due the LO-phonon resonances in a polar dielectric or due to the  interference effects in the substrates metalized on the back side. In some frequency intervals, the presence of the polar dielectric substrate compensates the strongly decreasing with $\omega$ frequency dependence of the third-order conductivity of graphene making the response almost frequency independent.
\end{abstract}


\maketitle

\tableofcontents

\section{Introduction}

It was theoretically predicted \cite{Mikhailov07e} and then experimentally confirmed \cite{Dragoman10,Hendry10} that, due to the linear energy dispersion of quasiparticles in graphene -- electrons and holes, this material should demonstrate strongly nonlinear electrodynamic properties. In recent years a great interest to the nonlinear electrodynamic phenomena in graphene arose, which was stimulated not only by purely academic interest but also by great perspectives which are opened up for using graphene in microwave-, terahertz- and opto-electronic devices. A variety of different nonlinear phenomena, such as harmonics generation\cite{Dragoman10,Bykov12,Kumar13,Hong13}, frequency mixing\cite{Hendry10,Hotopan11,Gu12}, saturable absorption\cite{Zhang12,Popa10,Popa11} and so on, have been already experimentally observed; even more have been theoretically predicted\cite{Mikhailov08a,Mikhailov09a,Mikhailov09b,Dean09,Dean10,Ishikawa10,Mikhailov11c,Jafari12,Mikhailov12c,Avetissian13,Mikhailov13c,Cheng14a,Cheng14b,Yao14,Smirnova14,Peres14,CoxAbajo14,CoxAbajo15,Savostianova15,Cheng15,Cheng16,Mikhailov16a,Mikhailov16b,Rostami16,CoxAbajo16,MariniAbajo16,Sharif16,Mikhailov16c}, for recent reviews see \cite{Glazov14,Hartmann14}. 

The higher harmonics generation under a monochromatic irradiation of a nonlinear medium is one of the fundamental nonlinear phenomena. In uniform graphene the second harmonic cannot be observed due to its central symmetry; therefore the lowest higher harmonic which can be emitted from uniform graphene is the third one. An analytical quantum theory of the third-order response of graphene has been recently developed in Refs. \cite{Cheng15,Cheng16,Mikhailov16a}. It was shown that the largest up-conversion efficiency ($\omega\to3\omega$) is achieved in the low-frequency (microwave, terahertz), quasi-classical regime $\hbar\omega\lesssim E_F$ where the inter-band electronic transitions between the valence and conduction bands of graphene can be neglected. At larger (infrared) frequencies $\hbar\omega\simeq E_F$ the effect is quantitatively smaller but a number of sharp and narrow resonances related to the three-photon ($3\hbar\omega= 2E_F$), two-photon ($2\hbar\omega= 2E_F$) and one-photon ($\hbar\omega= 2E_F$) transitions at the absorption edge have been predicted. The positions of these resonances depend on the electron/hole density in graphene ($E_F\propto\sqrt{n_s}$) and hence on the gate voltage, therefore the infrared resonances in the third-order response function of graphene are very interesting for a potential fine electric control of the third-harmonic generation.

The theory \cite{Cheng15,Cheng16,Mikhailov16a} was developed for a single, isolated graphene layer. In most experiments, however, graphene lies on a dielectric substrate which is or can be covered by metal on the back side. Recently we have investigated how the presence of a substrate may influence the third harmonic intensity $I_{3\omega}$ emitted from such a structure \cite{Savostianova15}. In particular, we have shown that the third harmonic intensity, emitted from graphene lying on a dielectric metalized from the back side, can be larger than that from an isolated graphene layer by more than two orders of magnitude. The physical reason of such a huge enhancement of the up-conversion efficiency is the interference of both the incident wave (with the frequency $\omega$) and the emitted wave ($3\omega$) in the dielectric, see a discussion in Ref. \cite{Savostianova15}. 

The results of Ref. \cite{Savostianova15} suppose that the dielectric constant $\epsilon$ (or the refractive index $n=\sqrt{\epsilon}$) of the dielectric substrate does not depend on the frequency, in particular, that $n_\omega=n_{3\omega}$. In real materials the frequency dispersion of the dielectric permittivity $\epsilon(\omega)$ is not negligible. On the other hand, the dielectric function $\epsilon(\omega)$ itself may have strong resonances, related, e.g., to the excitation of the transverse and longitudinal phonons in polar dielectrics (for example, in SiO$_2$, Al$_2$O$_3$). These resonances may also influence the intensity of the third harmonics emitted from graphene placed on (polar) dielectric and possibly metalized substrates.

In this paper we investigate in details the influence of different types of substrates on the third-harmonic generation from graphene. In Section \ref{sec:theory} we overview the applied theoretical approach. The results obtained are presented in Section \ref{sec:results}: we consider and analyze the structures AGA, AGDA, AGDM, AGPA, and AGPM, where the letters stand for air (A), graphene (G), dispersionless dielectric (D), polar dielectric (P), and metal (M). We show that the third harmonic intensity can be dramatically increased not only due to the interference of $\omega$- and $3\omega$-waves in the back-metalized structures (the effect partly considered in Ref. \cite{Savostianova15}) but also due to the LO-phonon resonances at the frequencies lying in the Reststrahlen-Band $\omega_{\rm TO}\lesssim \omega \lesssim \omega_{\rm LO}$ as well as above the LO-phonon frequency $\omega_{\rm LO}$. We also study the response of graphene on a substrate at low (terahertz) frequencies. In Section \ref{sec:summ} we summarize our results and draw conclusions. 

\section{Theoretical approach\label{sec:theory}}

\begin{figure}
\includegraphics[width=0.495\textwidth]{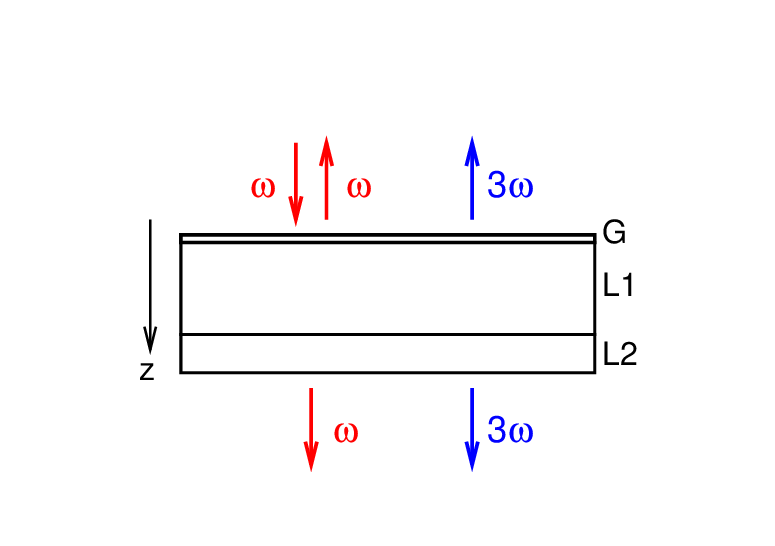}
\caption{\label{fig:geom} The geometry of the considered structure. A wave with the frequency $\omega$ is incident on the structure graphene (G) -- layer 1 (L1) -- layer 2 (L2) in the $z$ direction. The system emits the third harmonic $3\omega$ in the backward and forward directions. }
\end{figure}

We consider a structure shown in Figure \ref{fig:geom}. A linearly polarized (in the $x$-direction) electromagnetic wave with the frequency $\omega$ and intensity $I_\omega$ propagates in the positive $z$-direction and is incident on graphene (G) lying on a substrate. The substrate consists of two layers, L1 and L2, which can be dielectric or metallic. Graphene produces the third harmonic which is emitted in the positive (forward) and negative (backward) $z$-direction.

To find the intensity of the third harmonic we solve the system of Maxwell equations
\be 
\rot {\bm E}=-\frac {1}c\frac{\p {\bm H}}{\p t},\label{rotE}
\ee
\be 
\rot {\bm H}=\frac {\epsilon(\omega,z)}c\frac{\p {\bm E}}{\p t}+\frac{4\pi}c{\bm j}\delta(z),\label{rotH}
\ee
with conventional boundary conditions: the continuity of $E_x$ and $H_y$ fields at all boundaries except the boundary $z=0$ where the magnetic field has a jump
\be 
H_y(+0)-H_y(-0)=-\frac{4\pi}cj_x(0).\label{bc}
\ee
The dielectric permittivity $\epsilon(\omega,z)$ is constant (does not depend on $z$) inside each layer but may be a function of frequency. In the P-layer (polar dielectric) we assume that it has a single TO-phonon resonance so that
\be 
\epsilon_P(\omega)=\epsilon_\infty\left(1+\frac{\omega_{\rm LO}^2-\omega_{\rm TO}^2} {\omega_{\rm TO}^2-\omega^2 -i\omega\gamma_{\rm TO}}\right),
\label{dielfuncP}
\ee
where $\omega_{\rm TO}$ and $\omega_{\rm LO}$ are the frequencies of the transverse and longitudinal phonons and $\gamma_{\rm TO}$ is a phenomenological relaxation rate of TO phonons; $\epsilon_\infty$ is the dielectric constant at infinite frequencies. In the M-layer (metal) we describe the dielectric response by the Drude model
\be 
\epsilon_M(\omega)=1-\frac{\omega_p^2}{\omega(\omega+i\gamma_m)},
\ee
where $\omega_{p}$ is the plasma frequency in the metal and $\gamma_{m}=1/\tau_m$ ($\tau_m$) is the corresponding relaxation rate (time). The dielectric constants $\epsilon_A$ and $\epsilon_D$ of the A- and D-layers are frequency independent.

The current $\bm j$ in Eq. (\ref{rotH}) contains the first and the third harmonics,
\be 
j_x(t)=\sigma_{xx}^{(1)}(\omega)E_\omega^x(0)e^{-i\omega t} + \sigma_{xxxx}^{(3)}(\omega,\omega,\omega)\left[E_\omega^x(0)\right]^3e^{-i3\omega t} + \textrm{c.c.}
\ee
Here $E_\omega^x(0)$ is the complex amplitude of the electric field at the plane $z=0$, c.c. means the complex conjugate, and the functions $\sigma^{(1)}_{xx}(\omega)$ and $\sigma_{xxxx}^{(3)}(\omega,\omega,\omega)$ are the linear- and the third-order response conductivities of graphene. The explicit expression for $\sigma^{(1)}_{\alpha\beta}(\omega)$,
\be 
\sigma^{(1)}_{\alpha\beta}(\omega)=\delta_{\alpha\beta} \frac{e^2}{\pi\hbar}
\left(
\frac{i}{\Omega+i\Gamma}+\frac {i}{4}
\ln\frac{2-(\Omega + i \Gamma)}{2+(\Omega + i \Gamma)}\right),\label{sigma1}
\ee 
can be found, e.g. in Refs. \cite{Mikhailov07d,Mikhailov16a}; here $\Omega=\hbar\omega/E_F$, $\Gamma=\hbar\gamma/E_F=\hbar/\tau E_F$, $\gamma$ ($\tau$) is the phenomenological relaxation rate (time) of electrons in graphene. The explicit analytical expressions for  $\sigma_{\alpha\beta\gamma\delta}^{(3)}(\omega_1,\omega_2,\omega_3)$ can be found in Refs. \cite{Mikhailov16a,Cheng15,Cheng16}; they are very long [see, e.g. Eqs. (59)--(78) in Ref. \cite{Mikhailov16a}] and we do not reproduce them here. 

The problem is solved in two steps. First, we solve the linear response problem and calculate the transmission, reflection and absorption coefficients of the fundamental-harmonic wave (frequency $\omega$), as well as the electric field $E_\omega^x(0)$ at the plane $z=0$. Then we solve Maxwell equations for the $3\omega$-harmonic and calculate the intensities of the emitted third-harmonic radiation in the backward and forward directions. We present our results in terms of the parameter $\eta^{(3)}$, 
\be 
I_{3\omega}=\eta^{(3)} I_\omega^3. \label{eta}
\ee
which does not depend on the intensity of the incident radiation and describes the efficiency of the first- to third-harmonic transformation. The quantity $\eta^{(3)}$ is measured in units (cm$^2/W$)$^{2}$; the value of $\eta^{(3)}=10^{-18}$ (cm$^2/W$)$^{2}$ (see Figures below) means that 1 MW/cm$^2$ of the input ($\omega$) radiation produces 1 W/cm$^2$ of the output ($3\omega$) signal.

For specific results presented in the next Section \ref{sec:results} we use the following numerical parameters. In the dielectric layer we assume $\epsilon_D=4$ which is close, e.g., to the dielectric constant (3.9) of SiO$_2$. In polar dielectrics the typical values of the TO- and LO-phonon frequencies lie between $\sim 10$ and $30$ THz, see, e.g., Ref.\cite{Schubert00}. We assume therefore $\epsilon_\infty=1$, $\omega_{\rm TO}/2\pi=15$ THz, $\omega_{\rm LO}/2\pi=30$ THz, and $\gamma_{\rm TO}/2\pi=0.2$ THz. For the metallic layer we use parameters of gold \cite{Johnson72,Olmon12}: $\hbar\omega_p=8.45$ eV and $\tau_m=14$ fs. The dielectric constant of air is $\epsilon_A=1$. 

In the next Section we present results obtained this way for isolated graphene and different layered structures. 

\section{Results\label{sec:results}}

\subsection{Structure AGA \label{sec:aga}}

\begin{figure}
\includegraphics[width=0.495\textwidth]{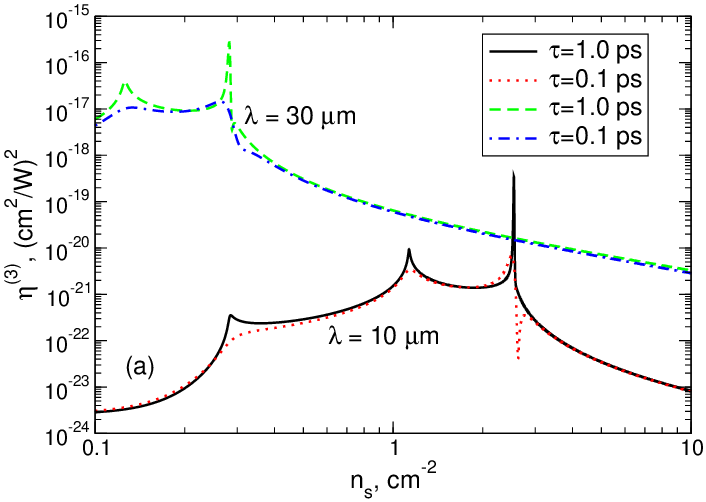}
\includegraphics[width=0.495\textwidth]{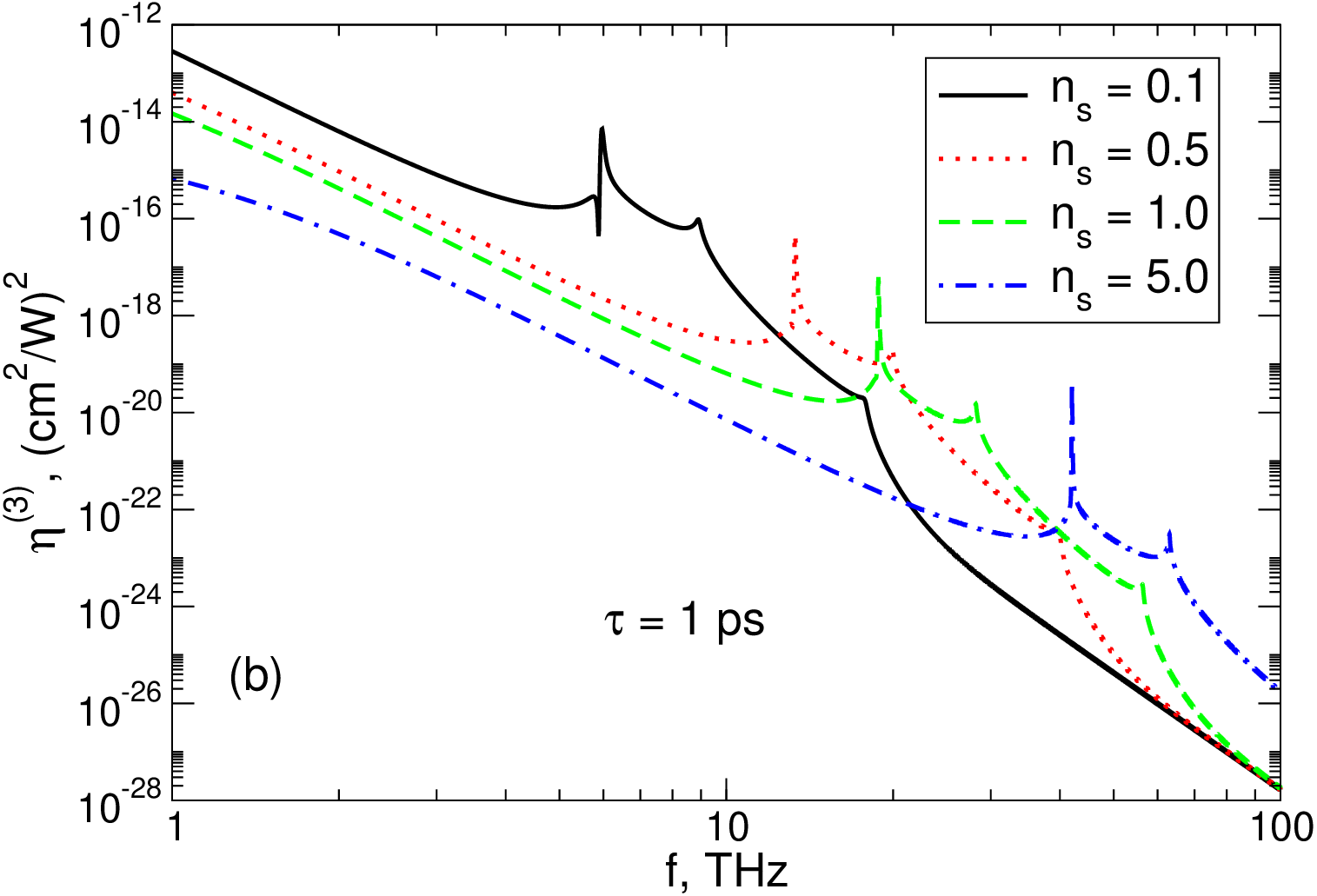}
\caption{\label{fig:AGA} The parameter $\eta^{(3)}$ of the AGA structure (a) as a function of the electron density at different input-wave frequencies ($f=30$ THz, $\lambda=10$ $\mu$m and $f=10$ THz, $\lambda=30$ $\mu$m) and different values of the relaxation time $\tau$ in graphene and (b) as a function of the input-wave frequency $f$ at $\tau=1$ ps and at different values of the electron density, measured in units $10^{12}$ cm$^{-2}$. The resonances correspond to $\hbar\omega/E_F=2/3$, $1$, and $2$; in (a) from right to left, in (b) from left to right.}
\end{figure}

We begin our analysis from the isolated graphene layer, i.e. the structure AGA (air -- graphene -- air). In this case the third-harmonic intensity depends on two dimensionless parameters, $\hbar\omega/E_F$ and $\hbar/\tau E_F$, Ref. \cite{Mikhailov16a}. Figure \ref{fig:AGA}(a) shows the parameter $\eta^{(3)}$ as a function of the electron density at two different values of the input-wave frequency and two values of the relaxation time $\tau$, compare with Fig. 7(a) from Ref. \cite{Mikhailov16a}. One sees that, first, the up-conversion coefficient $\eta^{(3)}$ strongly depends on the input-wave frequency: the increase of the input-wave wavelength $\lambda_\omega=2\pi c/\omega$ by a factor of 3 leads to the increase of $\eta^{(3)}$ by four orders of magnitude. Second, the density-dependence of $\eta^{(3)}$ contains three sharp resonances, at $\hbar\omega=2E_F/3$, $E_F$, and $2E_F$, which corresponds to the three-photon (the largest resonance), two-photon and one-photon (the smallest one) inter-band transitions at the absorption edge. Third, the overall behavior of the function $\eta^{(3)}(n_s)$ is not very sensitive to the relaxation time $\tau$, except the near-resonance regions. The resonances become very pronounced and sharp when the (inter-band) relaxation time gets bigger than $\sim 1$ ps: one sees, for example, that the increase of $\tau$ by one order of magnitude leads to a more than (almost) two orders of magnitude increase of $\eta^{(3)}$ at $\lambda_\omega=10$ $\mu$m (30 $\mu$m).  

Figure \ref{fig:AGA}(b) shows the up-conversion coefficient as a function of frequency at a few values of $n_s$ and at $\tau=1$ ps. One sees a strong reduction of the effect with $\omega$ and sharp resonances related to the inter-band multi-photon transitions. The coefficient $\eta^{(3)}$ can be of order $10^{-13}-10^{-12}$ (cm$^2/W$)$^{2}$ at the input-wave frequency $f\simeq 1$ THz and of order $10^{-18}-10^{-17}$ (cm$^2/W$)$^{2}$ (at low densities and in out-of-resonance regions) at the input-wave frequency $f\simeq 10$ THz. Using the $2E_F/3$ resonances one can increase the last values by about two orders of magnitude if $\tau=1$ ps.

\subsection{Structure AGDA \label{sec:agda}}

\begin{figure}
\includegraphics[width=0.495\textwidth]{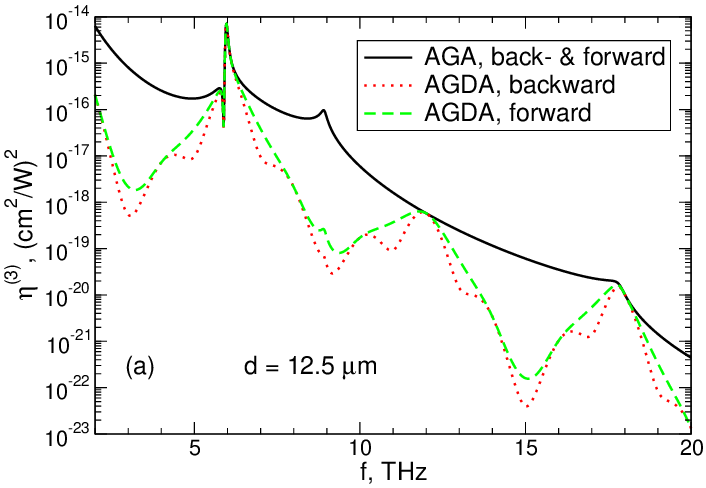}
\includegraphics[width=0.495\textwidth]{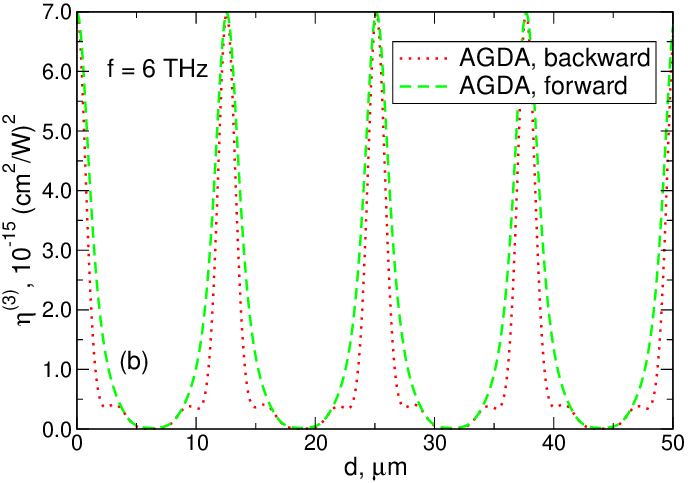}
\caption{\label{fig:AGDA} The parameter $\eta^{(3)}$ of the AGDA structure (a) as a function of the input-wave frequency $f$ at the dielectric thickness $d=12.5$ $\mu$m and (b) as a function of $d$ at $f=6$ THz ($\lambda=50$ $\mu$m). Other parameters are $\tau=1$ ps and $n_s=10^{11}$ cm$^{-2}$; the refractive index of the dielectric is 2. The black curve in (a) shows for comparison the result for the isolated graphene layer (AGA) corresponding to the black solid curve in Fig. \ref{fig:AGA}(b). Red dotted and green dashed curves show the third-harmonic wave intensity emitted in the backward and forward direction respectively. }
\end{figure}

Now consider graphene lying on the dielectric substrate with the refractive index $n_\omega=n_{3\omega}=2$ and a dielectric thickness $d$ (the structure AGDA). This situation was preliminary considered in Ref. \cite{Savostianova15}; here we analyze it in more details, in particular, as a function of the input-wave frequency. We also need the results for the AGDA structure for the sake of comparison with the polar dielectric case (Section \ref{sec:agpa}). Figure \ref{fig:AGDA}(a) shows the up-conversion efficiency $\eta^{(3)}$ as a function of the input-wave frequency at the electron density $n_s=10^{11}$ cm$^{-2}$ and $d=12.5$ $\mu$m. First, one sees that the third-harmonic intensity from the AGDA system is always smaller than from the isolated graphene layer (AGA). At several values of the frequency ($\simeq 6$, 12, 18 THz) the value of $\eta^{(3)}$ for the AGDA structure coincides with that for the AGA structure. This is a consequence of the interference of both the first and third harmonics in the dielectric slab and these frequencies satisfy the conditions
\be 
d=\frac{\lambda_\omega}{2n_\omega}m_1, \ \ \ m_1=0,1,2,\dots,   \label{interfer-cond1}
\ee
and
\be 
d=\frac{\lambda_{3\omega}}{2n_{3\omega}}m_2, \ \ \ m_2=0,1,2,\dots,  \label{interfer-cond3} 
\ee
i.e. the slab thickness $d$ is a multiple of the half-wavelength of the first and third harmonics. Since we assume for the dielectric (D) that $n_{\omega}=n_{3\omega}=const$, the second set of conditions (\ref{interfer-cond3}) is a subset of the first one (\ref{interfer-cond1}), so that the points where $\eta^{(3)}_{\rm AGDA}=\eta^{(3)}_{\rm AGA}$ are determined by the condition (\ref{interfer-cond1}). 

Second, the intensity of the third harmonic emitted in the forward direction is always higher than in the backward direction. The factor $\eta^{(3)}_{\rm AGDA}$ for the third harmonic emitted back has additional oscillations with the local maxima determined by Eq. (\ref{interfer-cond3}) (the interference of the third harmonic wave). These interference features are also seen in Figure \ref{fig:AGDA}(b) where we show the factor $\eta^{(3)}$ as a function of $d$ at fixed values of $\tau$, $n_s$ and $f$. 

In Figure \ref{fig:AGDA}(a) we have chosen the electron density ($n_s\approx 10^{11}$ cm$^{-2}$) and the dielectric thickness ($d=12.5$ $\mu$m) so that one of the interference maxima (\ref{interfer-cond1}), namely, the one with $m_1=1$, coincided with the largest inter-band resonance at
\be 
\omega=\omega_{\rm res}=\frac{2E_F}{3\hbar}=\frac{2 }{3}v_F\sqrt{\pi n_s};\label{Wres}
\ee
otherwise, the large $\eta^{(3)}$-value ($\simeq 7\times 10^{-15}$ cm$^4$/W$^2$)
would be strongly suppressed by the destructive interference of waves reflected from the dielectric surfaces. The optimal condition for the observation of a resonant enhancement of the third harmonic in the AGDA structure is thus
\be 
\omega=\frac{2 }{3}v_F\sqrt{\pi n_s}=\frac{\pi c}{nd}m_1,
\label{opt-cond}
\ee
where $n=n_\omega=n_{3\omega}$ is assumed to be constant. The condition (\ref{opt-cond}) relates the frequency, the electron density and the thickness of the dielectric slab. 

The number $\eta^{(3)}\simeq 7\times 10^{-15}$ cm$^4$/W$^2$ in our example corresponds to the emission of $7$ W/cm$^2$ of the third-harmonic (18 THz) at the input wave (6 THz) intensity of 100 kW/cm$^2$.

\subsection{Structure AGDMA \label{sec:agdma}}

\begin{figure}
\includegraphics[width=0.495\textwidth]{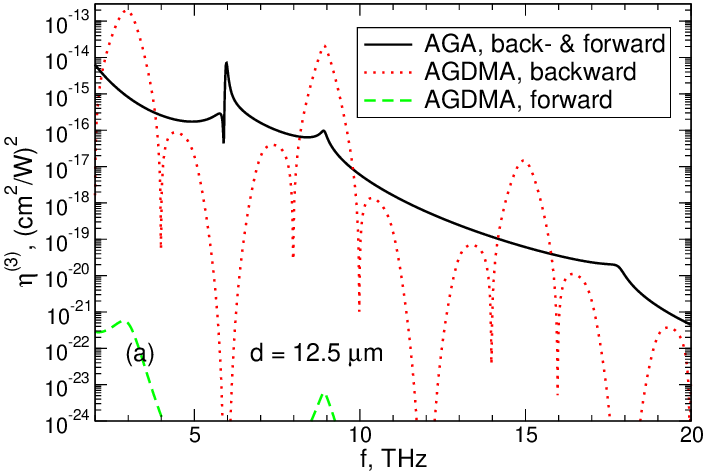}
\includegraphics[width=0.495\textwidth]{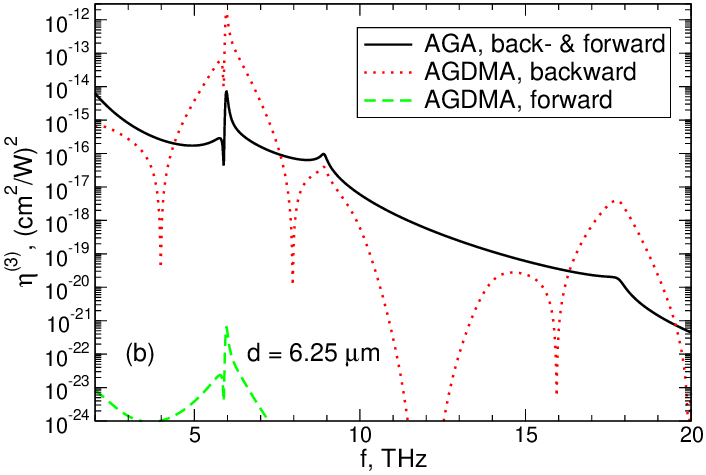}
\caption{\label{fig:AGDM} The parameter $\eta^{(3)}$ of the AGDMA structure as a function of the input-wave frequency $f$ at the dielectric thickness of (a) $d=12.5$ $\mu$m and (b) $d=6.25$ $\mu$m. Other parameters are $\tau=1$ ps, $n_s=10^{11}$ cm$^{-2}$; the refractive index of the dielectric is 2, the metal (Au) thickness is 0.2 $\mu$m. The black curve in (a) shows for comparison the result for the isolated graphene layer (AGA) corresponding to the black solid curve in Fig. \ref{fig:AGA}(b). Red dotted and green dashed curves show the third-harmonic wave intensity (at the frequency $3f$) emitted in the backward and forward direction respectively.}
\end{figure}

Now consider what happens if the same structure that was analyzed in Section \ref{sec:agda} is covered by a thin metallic (Au) layer on the back side. Figure \ref{fig:AGDM}(a) shows the dependence of $\eta^{(3)}$ on the input-wave frequency at the same parameters as in Figure \ref{fig:AGDA}(a), in particular, at $d=12.5$ $\mu$m; the only difference is that the backside of the dielectric is covered by a thin (0.2 $\mu$m) layer of gold. One sees that the spectrum of the up-conversion coefficient $\eta^{(3)}$ dramatically changes. First, since the metal reflects both the first and third harmonics, the emission of the $3\omega$-wave in the forward direction is suppressed by many orders of magnitude and can be completely neglected. Second, the emission of the $3\omega$-wave in the backward direction is strongly modified, and in certain frequency ranges, e.g. around the input-wave frequency $f=3$, 9, and 15 THz the coefficient $\eta^{(3)}$ is substantially (by orders of magnitude) larger than in the AGA structure. For example, at $f=3$ THz the value of $\eta^{(3)}_{\rm AGDMA}$ is about $2\times 10^{-13}$ cm$^4$/W$^2$, while $\eta^{(3)}_{\rm AGA}\simeq 8\times 10^{-16}$ cm$^4$/W$^2$ (the growth by a factor of $\sim 250$). At $f=9$ THz (which corresponds to the resonance $\hbar\omega=E_F$) the corresponding numbers are $\eta^{(3)}_{\rm AGDMA}\simeq 2.1\times 10^{-14}$ cm$^4$/W$^2$ and $\eta^{(3)}_{\rm AGA}\simeq 9.8\times 10^{-17}$ cm$^4$/W$^2$ (the growth by a factor of $\sim 210$). At $f=15$ THz we have $\eta^{(3)}_{\rm AGDMA}\simeq 1.5\times 10^{-17}$ cm$^4$/W$^2$ and $\eta^{(3)}_{\rm AGA}\simeq 6.2\times 10^{-20}$ cm$^4$/W$^2$ (the growth by a factor of $\sim 240$). 

On the other hand, the resonance at $\hbar\omega=2E_F/3$ ($\simeq 6$ THz), which is the largest inter-band resonance in the AGA structure, is completely suppressed (by more than ten orders of magnitude). This is explained, again, by the interference of waves in the dielectric, see \cite{Savostianova15}. In the presence of metal on its back side, however, the boundary condition at $z=d$ (the boundary dielectric -- metal) changes: the tangential electric field $E_x$ should vanish at this point and the interference maxima in the AGDMA structure are expected at 
\be 
d=\frac{\lambda_\omega}{2n_\omega}\left(m_1+\frac 12\right), \ \ \ m_1=0,1,2,\dots ,   \label{interfer-cond-M}
\ee
i.e. the dielectric thickness is a quarter of wavelength plus an integer times a half-wavelength. In Figure \ref{fig:AGDM}(b) we plot the dependence of $\eta^{(3)}_{\rm AGDMA}$ on the input-wave frequency $f$ at two times smaller dielectric thickness which corresponds to the condition (\ref{interfer-cond-M}) with $m_1=0$. The resonances at 3 and 9 THz are suppressed while those at 6 THz and 18 THz (correspond to the $\hbar\omega=2E_F/3$ and $2E_F$ resonances) become substantially stronger and exceed the corresponding values of $\eta^{(3)}_{\rm AGA}$. In particular, at $f=6$ THz the value of $\eta^{(3)}_{\rm AGDMA}$ is about $1.7\times 10^{-12}$ cm$^4$/W$^2$, as compared to $\eta^{(3)}_{\rm AGA}\simeq 7\times 10^{-15}$ cm$^4$/W$^2$ (the growth by a factor of $\sim 240$). At $f=18$ THz we have, similarly, $\eta^{(3)}_{\rm AGDMA}\simeq 4\times 10^{-18}$ cm$^4$/W$^2$ and $\eta^{(3)}_{\rm AGA}\simeq 1.9\times 10^{-20}$ cm$^4$/W$^2$ (the growth by a factor of $\sim 210$). 

The value of $\eta^{(3)}\simeq 1.7\times 10^{-12}$ cm$^4$/W$^2$, which can be achieved under the chosen parameters at the incident wave frequency of 6 THz, corresponds to the emission of $1.7$ kW/cm$^2$ of the third-harmonic (18 THz) at the input wave intensity of 100 kW/cm$^2$, compare with the last sentence in Section \ref{sec:agda}. Or, this means that the third-harmonic intensity of $1.7$ W/cm$^2$ can be achieved already at the quite low intensity of the incident wave of only 10 kW/cm$^2$.

\subsection{Structure AGPA \label{sec:agpa}}

\begin{figure}
\includegraphics[width=0.495\textwidth]{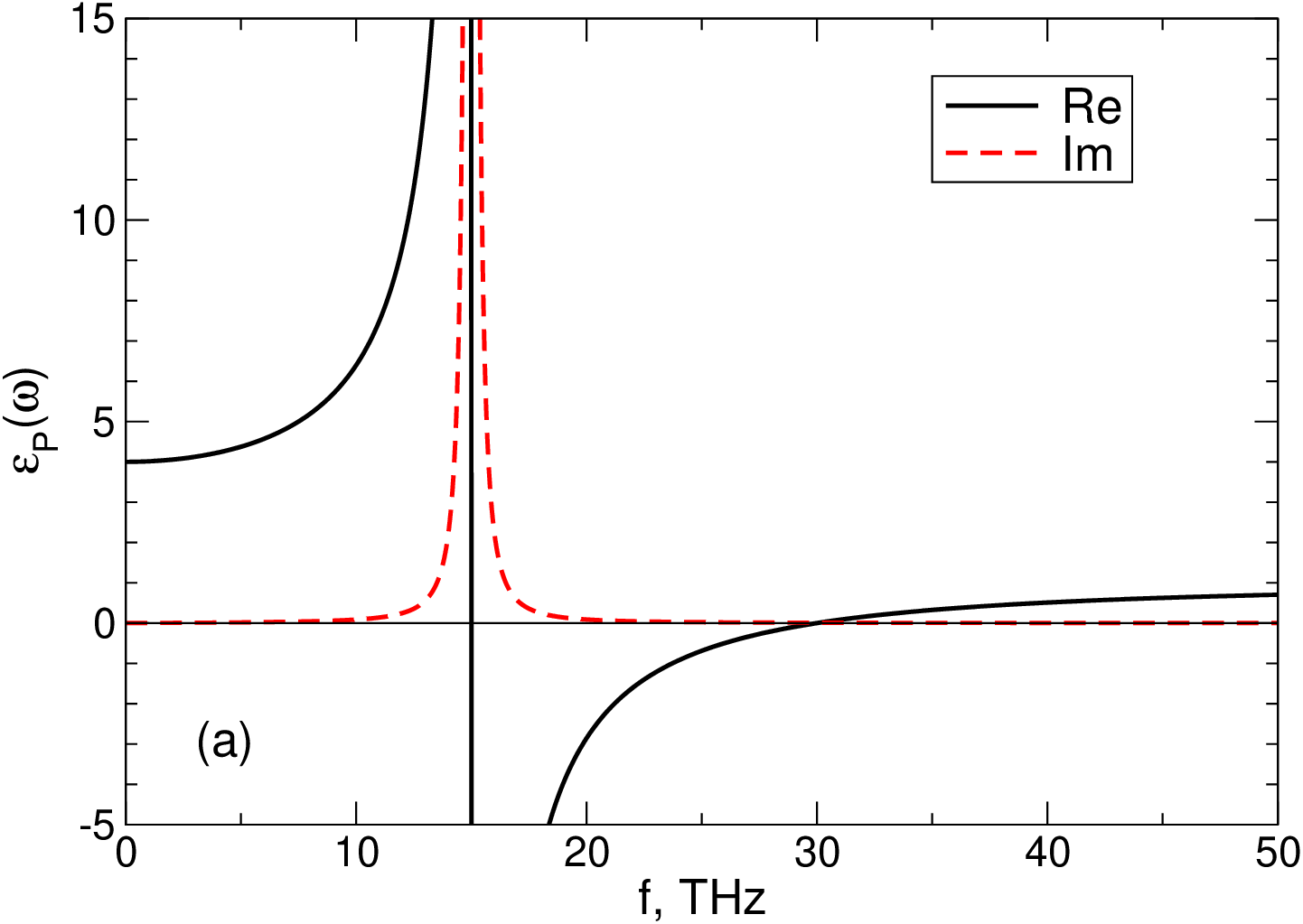}
\includegraphics[width=0.495\textwidth]{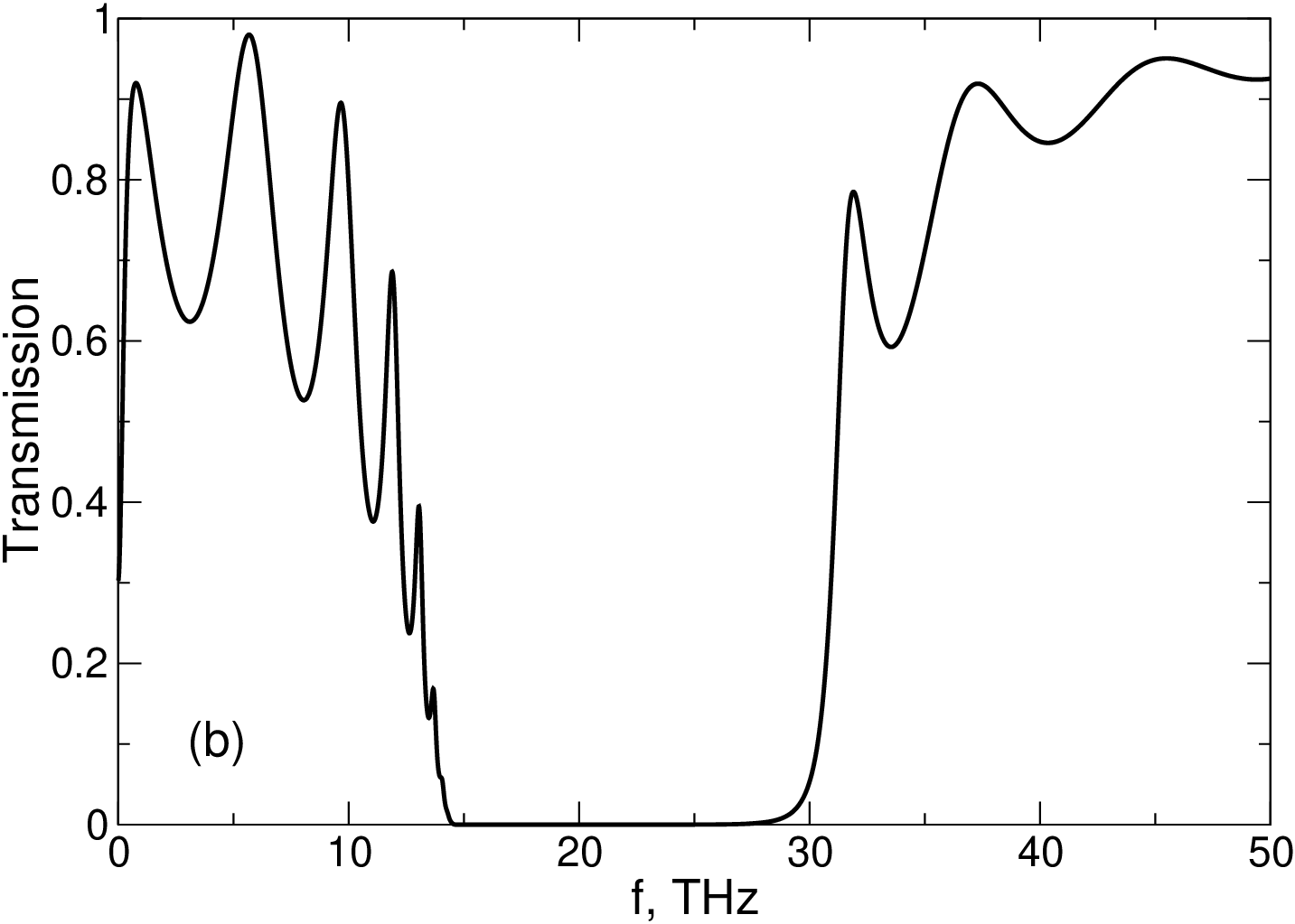}
\includegraphics[width=0.495\textwidth]{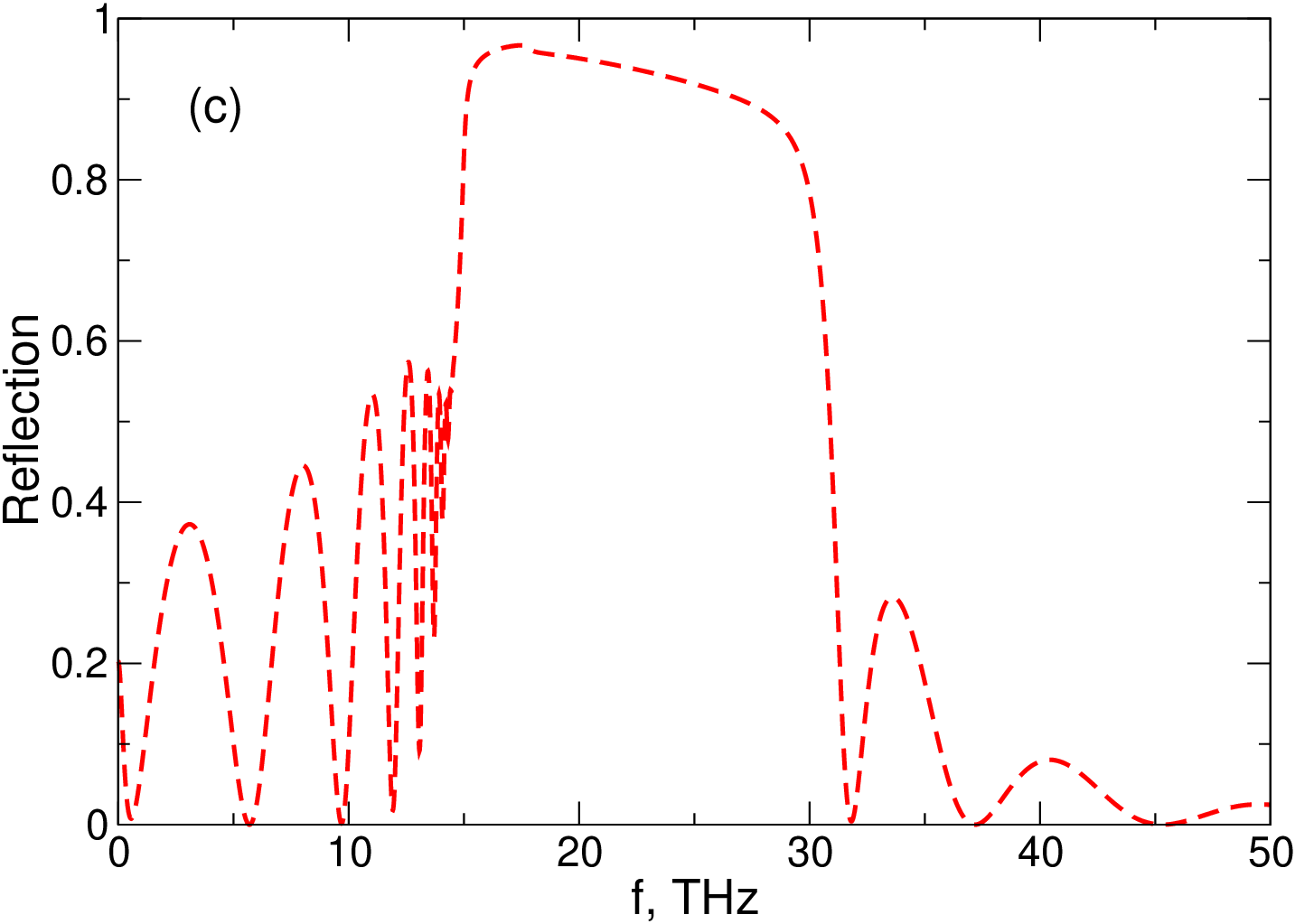}
\includegraphics[width=0.495\textwidth]{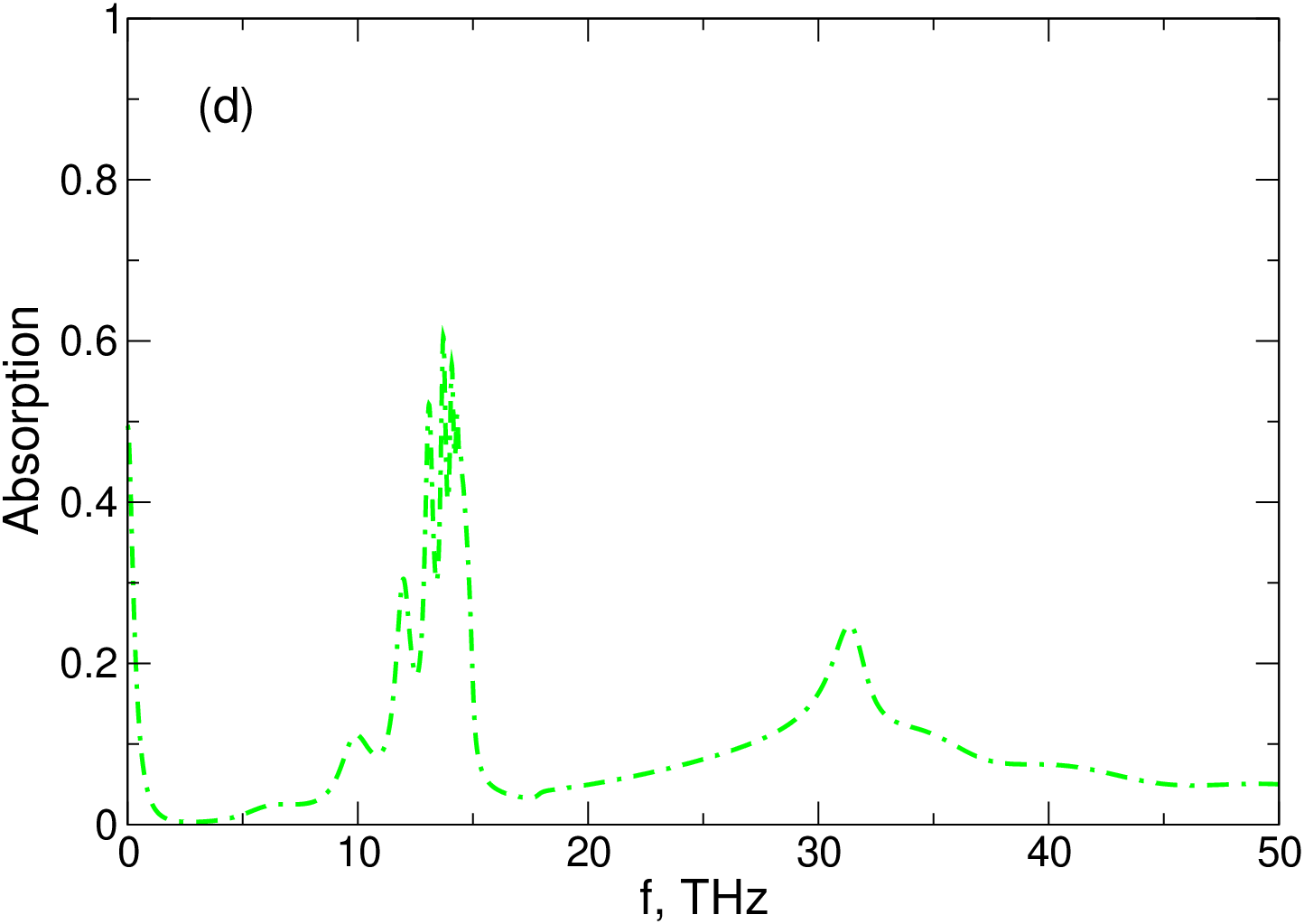}
\caption{\label{fig:AGPA-lin} (a) The real and imaginary parts of the dielectric function $\epsilon_P(\omega)$, Eq. (\ref{dielfuncP}), as well as (b) the transmission, (c) the reflection and (d) the absorption coefficients of an AGPA structure as a function of the input-wave frequency $f$ at the dielectric thickness $d=12.5$ $\mu$m. Parameters of the polar dielectric $\epsilon_\infty=1$, $\omega_{\rm TO}/2\pi=15$ THz, $\omega_{\rm LO}/2\pi=30$ THz, $\gamma_{\rm TO}/2\pi=0.2$ THz. Parameters of graphene are $\tau=1$ ps, $n_s=10^{11}$ cm$^{-2}$.}
\end{figure}

The dielectric substrate with a dispersionless dielectric permittivity considered in the previous Section is not very common. Usually the dielectric function is frequency-dependent in the mid-IR range which is related with the optical-phonons poles. Real polar dielectrics typically have several TO-phonon poles, but in order to clarify the physics of the discussed phenomena we restrict ourselves here by the model (\ref{dielfuncP}) with a single TO-phonon pole, Figure \ref{fig:AGPA-lin}(a). 

The third-order response of the AGPA structure is much more complicated, as compared to the cases considered in the previous Sections, therefore it makes sense to consider first the linear response spectra. Figure \ref{fig:AGPA-lin}(b)--(d) shows the transmission, reflection and absorption coefficients of the AGPA structure at $d=12.5$ $\mu$m (the same dielectric thickness as in Figures \ref{fig:AGDA}(a) and \ref{fig:AGDM}(a)). The TO-phonon frequency is chosen to lie at 15 THz and the LO-phonon frequency -- at 30 THz. Since $\omega_{\rm LO}/\omega_{\rm TO}=2$ and $\epsilon_\infty=1$, the static dielectric constant 
\be 
\epsilon_0=\epsilon_\infty\frac{\omega_{\rm LO}^2}{\omega_{\rm TO}^2}
\ee
equals 4 which coincides with $\epsilon_D$ in the dielectric layer considered in Sections \ref{sec:agda}--\ref{sec:agdma}. As seen from Figures \ref{fig:AGPA-lin}(b)--(d), in the frequency window from $\omega_{\rm TO}$ to $\omega_{\rm LO}$ the AGPA structure reflects almost all the radiation and the transmission coefficient is close to zero (the Reststrahlen-Band effect). In the areas $\omega\lesssim \omega_{\rm TO}$ and $\omega\gtrsim \omega_{\rm LO}$ the TRA-coefficients strongly oscillate which is related to the interference of the incident ($\omega$) radiation in the dielectric slab, according to Eq. (\ref{interfer-cond1}). In this equation now, the refractive index $n_\omega=\sqrt{\epsilon_P(\omega)}$  depends on the frequency (strongly increases when $\omega\to\omega_{\rm TO}$ and tends to zero when $\omega_{\rm LO}\leftarrow\omega$), therefore the oscillation periods are not constant but vary approaching the TO-phonon frequency from the left and the LO-phonon frequency from the right. The absorption coefficient has two main maxima at the frequency slightly lower than $\omega_{\rm TO}$ and slightly higher than $\omega_{\rm LO}$, with interference related oscillations. 

Now consider the third-order response. Having the goal to maximize the intensity of the third harmonic signal we can try to take advantage of combining the graphene resonance (\ref{Wres}) with the optical phonon resonances at $\omega_{\rm TO}$ or $\omega_{\rm LO}$. Therefore we consider four special cases. 

\subsubsection{The case $3\omega_{\rm res}\simeq \omega_{\rm TO}$\label{3resTO}}

\begin{figure}
\includegraphics[width=0.495\textwidth]{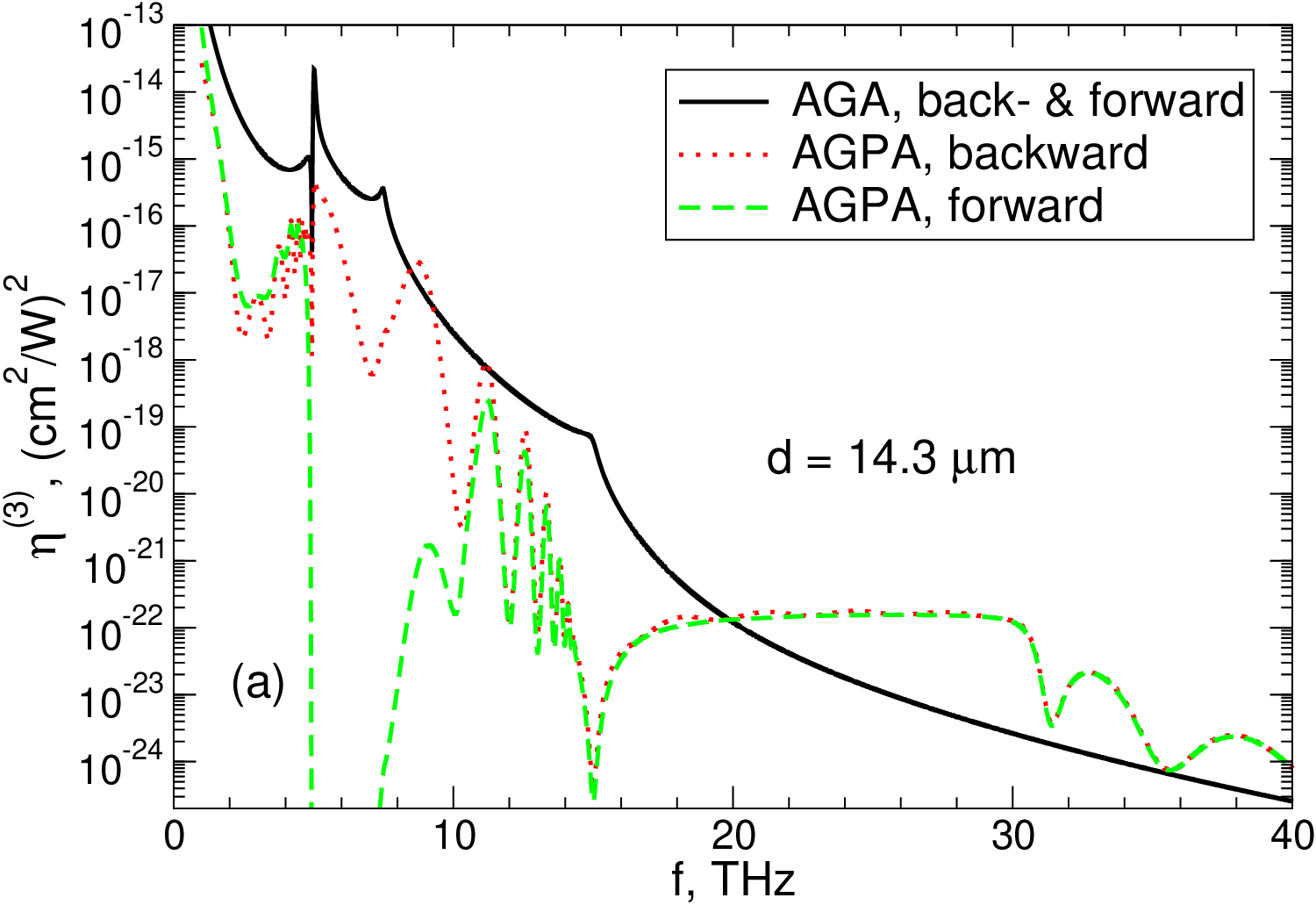}
\includegraphics[width=0.495\textwidth]{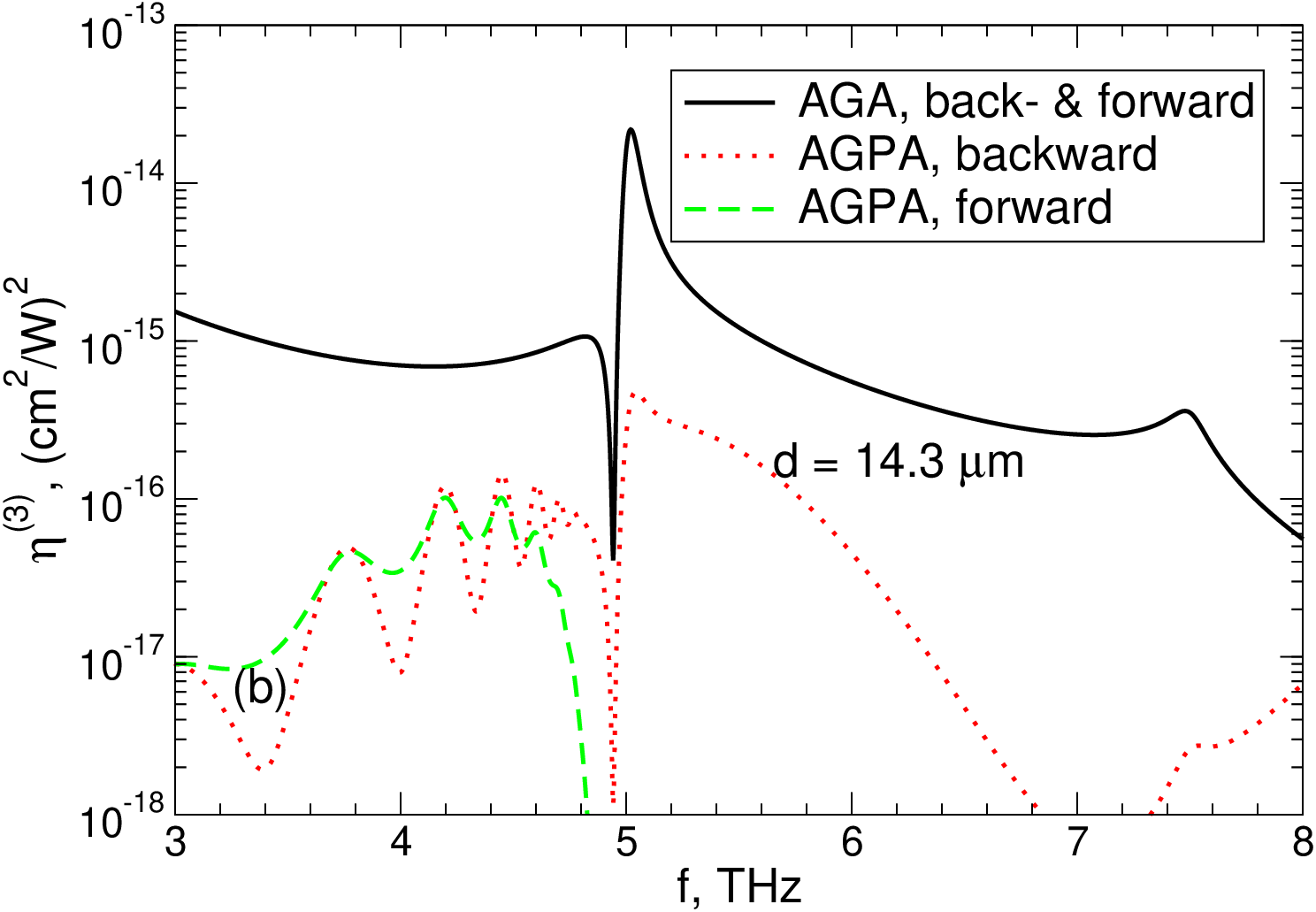}
\caption{\label{fig:AGPA-1a} The parameter $\eta^{(3)}$ of the AGPA structure as a function of the input-wave frequency $f$ at the polar dielectric thickness $d=14.3$ $\mu$m in (a) a broad range $0-40$ THz and (b) in the narrow frequency range around the resonance $\omega=\omega_{\rm res}$, Eq. (\ref{Wres}). Parameters of graphene are $\tau=1$ ps and $n_s=0.707\times 10^{11}$ cm$^{-2}$, parameters of the polar dielectric: $\epsilon_\infty=1$, $f_{\rm TO}=15$ THz, $f_{\rm LO}=30$ THz, $\gamma_{\rm TO}/2\pi=0.2$ THz. The black curves show for comparison the result for the isolated graphene layer (AGA). The red dotted and green dashed curves show the third-harmonic wave intensity (at the frequency $3f$) emitted in the backward and forward direction respectively.}
\end{figure}

\begin{figure}
\includegraphics[width=0.495\textwidth]{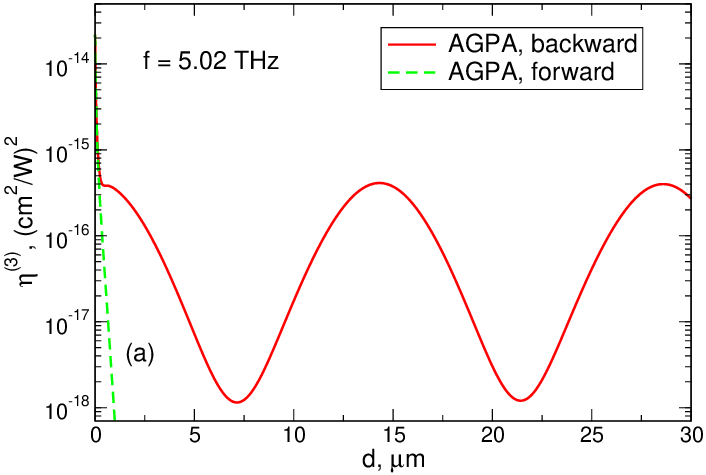}
\includegraphics[width=0.495\textwidth]{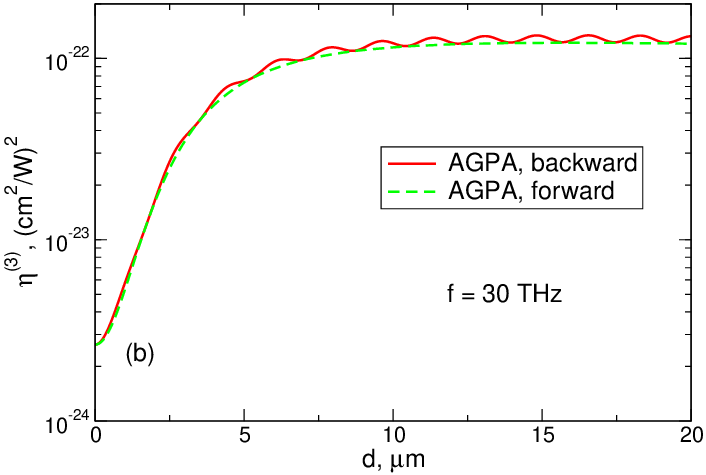}
\caption{\label{fig:AGPA-1b} The parameter $\eta^{(3)}$ of the AGPA structure  as a function of the dielectric thickness $d$ at the input-wave frequency of (a) $f=5.02$ THz (corresponds to the maximum of $\eta^{(3)}$ at the $\hbar\omega/E_F=2/3$ resonance) and (b) $f=30$ THz, away from the graphene resonance at the upper boundary of the Reststrahlen-Band. Parameters of graphene are $\tau=1$ ps and $n_s=0.707\times 10^{11}$ cm$^{-2}$, parameters of the polar dielectric: $\epsilon_\infty=1$, $f_{\rm TO}=15$ THz, $f_{\rm LO}=30$ THz, $\gamma_{\rm TO}/2\pi=0.2$ THz.  Red solid and green dashed curves show the third-harmonic wave intensity emitted in the backward and forward direction respectively. The value of the efficiency $\eta^{(3)}_{\rm AGPA}$ at $d\to 0$ in Figure (a), which is not clearly seen in the figure, is $\simeq 2.18\times 10^{-14}$ (cm$^2$/W)$^2$ both for the forward and backward emitted radiation.}
\end{figure}

Figures \ref{fig:AGPA-1a}(a,b) show the frequency dependence of $\eta^{(3)}$ for the AGA structure at $\tau=1$ ps, $n_s=7.06858\times 10^{10}$ cm$^{-2}$ and the AGPA structure at the same values of $\tau$ and $n_s$, and at $d=14.3$ $\mu$m. The density $n_s$ is chosen so that the graphene resonance (\ref{Wres}) lies at $f_{\rm res}=\omega_{\rm res}/2\pi\simeq 5$ THz; then $3f_{\rm res}$ coincides with the TO-phonon frequency 15 THz. The chosen thickness $d=14.3$ $\mu$m corresponds to one of the interference maxima, Figure \ref{fig:AGPA-1b}(a), of the coefficient $\eta^{(3)}$ at the incident wave frequency $f=5.02$ THz (the exact position of the resonance maximum in Figures \ref{fig:AGPA-1a}(a,b)). Several interesting features are seen in Figures \ref{fig:AGPA-1a}(a,b). First, the third-harmonic intensities, emitted in the forward and backward directions, are quite close to each other at $f\gtrsim 10$ THz, but are substantially different at the lower frequencies, especially at $5\lesssim f\lesssim 10$ THz where the forward-emitted $3\omega$-intensity falls down by many orders of magnitude. This is explained by the fact that at $5\lesssim f\lesssim 10$ THz the frequency $3f$ lies in the Reststrahlen-Band, $15\lesssim 3f\lesssim 30$ THz, therefore the $3\omega$-wave produced by graphene does not penetrate into the polar dielectric and can therefore be emitted only in the backward direction. Second, the oscillations of $\eta^{(3)}$ at $10\lesssim f\lesssim 15$ THz are due to the interference of the input-wave frequency harmonic in the dielectric: the oscillation period becomes shorter when approaching the TO-phonon frequency from the left, similar to the behavior of TRA coefficients discussed above, Figure \ref{fig:AGPA-lin}(b)--(d). It is interesting that a similar oscillating behavior of $\eta^{(3)}$ is also seen at the frequencies below $5$ THz, for a detailed picture see Figure \ref{fig:AGPA-1a}(b). This is due to the same interference effect but of the third harmonic: when $f$ is approaching 5 THz, its third harmonic tends to 15 THz which is just the TO-phonon pole where the refractive index $n_{3\omega}$ diverges. 

The influence of the polar dielectric on the third-harmonic intensity at the frequency $f=f_{\rm res}\simeq 5$ THz is seen in the detailed picture Figure \ref{fig:AGPA-1a}(b) as well as in Figure \ref{fig:AGPA-1b}(a). Although the thickness $d=14.3$ $\mu$m was chosen to maximize the third order response, Figure \ref{fig:AGPA-1b}(a), the coefficient  $\eta^{(3)}_{\rm AGPA}$ is by a factor of $\sim 50$ smaller than $\eta^{(3)}_{\rm AGA}$ at $f= 5.02$ THz. Figure \ref{fig:AGPA-1b}(a) shows how the coefficient $\eta^{(3)}_{\rm AGPA}$ varies with $d$: first, when $d$ grows from zero up to $\sim 0.4$ $\mu$m, the factor $\eta^{(3)}_{\rm AGPA}$ drastically drops down and then begins to oscillate due to the interference of the incident wave in the dielectric, Eq. (\ref{interfer-cond1}). Thus, if $3\omega_{\rm res}\simeq \omega_{\rm TO}$ the use of the polar dielectric does not help to increase the third-harmonic response at $f\simeq f_{\rm res}$.

However, at frequencies far from the resonance (\ref{Wres}), at $f\gtrsim 20$ THz in Figure \ref{fig:AGPA-1a}(a), the polar-dielectric substrate does help to substantially increase the third-harmonic generation. For example, at the incident-wave frequency $f\simeq 30$ THz the coefficient $\eta^{(3)}_{\rm AGA}$ of a pure graphene is about $2.65\times 10^{-24}$ (cm$^2$/W)$^2$, but the one of the AGPA structure, $\eta^{(3)}_{\rm AGPA}$, equals $1.29\times 10^{-22}$ (cm$^2$/W)$^2$, which is almost 50 times bigger. The growth of $\eta^{(3)}_{\rm AGPA}$ with the dielectric thickness $d$ is illustrated in Figure \ref{fig:AGPA-1b}(b). One sees that $\eta^{(3)}_{\rm AGPA}$ substantially grows when $d$ varies from zero up to $\approx 5$ $\mu$m; then it essentially saturates. The weak oscillations of the backward emitted third-harmonic intensity (the red curve in Figure \ref{fig:AGPA-1b}(b)) is the consequence of the interference of the third-harmonic wave (90 THz in our example) in the dielectric slab: at $3f=90$ THz the polar dielectric has the refractive index slightly smaller than one, and the oscillation period in Figure \ref{fig:AGPA-1b}(b) is well described by the interference formula (\ref{interfer-cond3}). 

\begin{figure}
\includegraphics[width=0.495\textwidth]{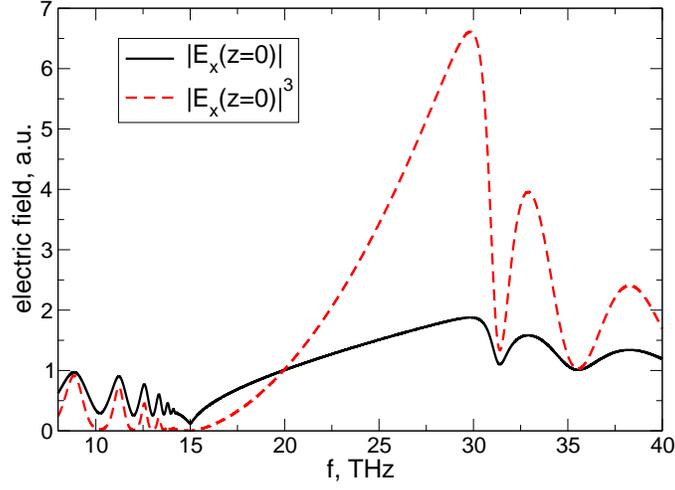}
\caption{\label{fig:field} The absolute value of the $\omega$-component of the electric field at the plane $z=0$, $|E_x(z=0)|$, and its cube as a function of the input-wave frequency $f$ in the vicinity of the Reststrahlen-Band $15-30$ THz. Parameters of the AGPA structure are the same as in Figure \ref{fig:AGPA-1a}.}
\end{figure}

Even more interesting feature seen in Figure \ref{fig:AGPA-1a}(a) is that the third-harmonic intensity very weakly depends on the input-wave frequency $f$ if it lies in the Reststrahlen-Band, $15\lesssim f\lesssim 30$ THz (the third-harmonic frequency $3f$ then lies between $\sim 45$ and 90 THz). This fact can be very interesting for application. Indeed, the third-order conductivity of graphene $\sigma_{xxxx}(\omega,\omega,\omega)$, responsible for the third-harmonic generation effect, falls down very quickly with $\omega$. Therefore the coefficient $\eta^{(3)}_{\rm AGA}$ of the isolated graphene quickly decreases with the frequency (by four orders of magnitude when $f$ varies from 15 THz to 30 THz, see Figure \ref{fig:AGPA-1a}(a)). The factor $\eta^{(3)}_{\rm AGPA}$ is almost frequency independent in this frequency range which means that the decrease of $\sigma_{xxxx}(\omega,\omega,\omega)$ should be compensated by something else. To understand the reason of this effect we plot the frequency dependence of the electric field $E_x(z=0)$ at the plane $z=0$, as well as of the cube of this field, see Figure \ref{fig:field}. One sees that the field grows in the interval from $f_{\rm TO}$ to $f_{\rm LO}$, even stronger grows the cube of the field; therefore, the frequency dependence of the third-order conductivity is to a large extent compensated. This explains the almost flat frequency dependence of the third-harmonic intensity in the range $3f_{\rm TO}\lesssim 3f\lesssim 3f_{\rm LO}$ 
when the input-wave frequency lies in the Reststrahlen-Band $f_{\rm TO}\lesssim f\lesssim f_{\rm LO}$. 

When the incident wave frequency exceeds $f_{\rm LO}$, the coefficient $\eta^{(3)}_{\rm AGPA}$ begins to oscillate, again, due to the interference of the input-wave frequency wave in the dielectric slab. The points where $\eta^{(3)}_{\rm AGPA}$ decreases and almost touches $\eta^{(3)}_{\rm AGA}$, Figure \ref{fig:AGPA-1a}(a), i.e. approximately at 31.5, 35.5 and 42.1 THz (the last point is not shown in the Figure) correspond to the interference oscillations described by Eq. (\ref{interfer-cond1}) with $m_1=1,2$ and 3 respectively; the growing with $m_1$ period of oscillations is explained by the frequency dependence of the refractive index $n_\omega$.

\subsubsection{The case $\omega_{\rm res}\simeq \omega_{\rm TO}$\label{resTO}}

\begin{figure}
\includegraphics[width=0.495\textwidth]{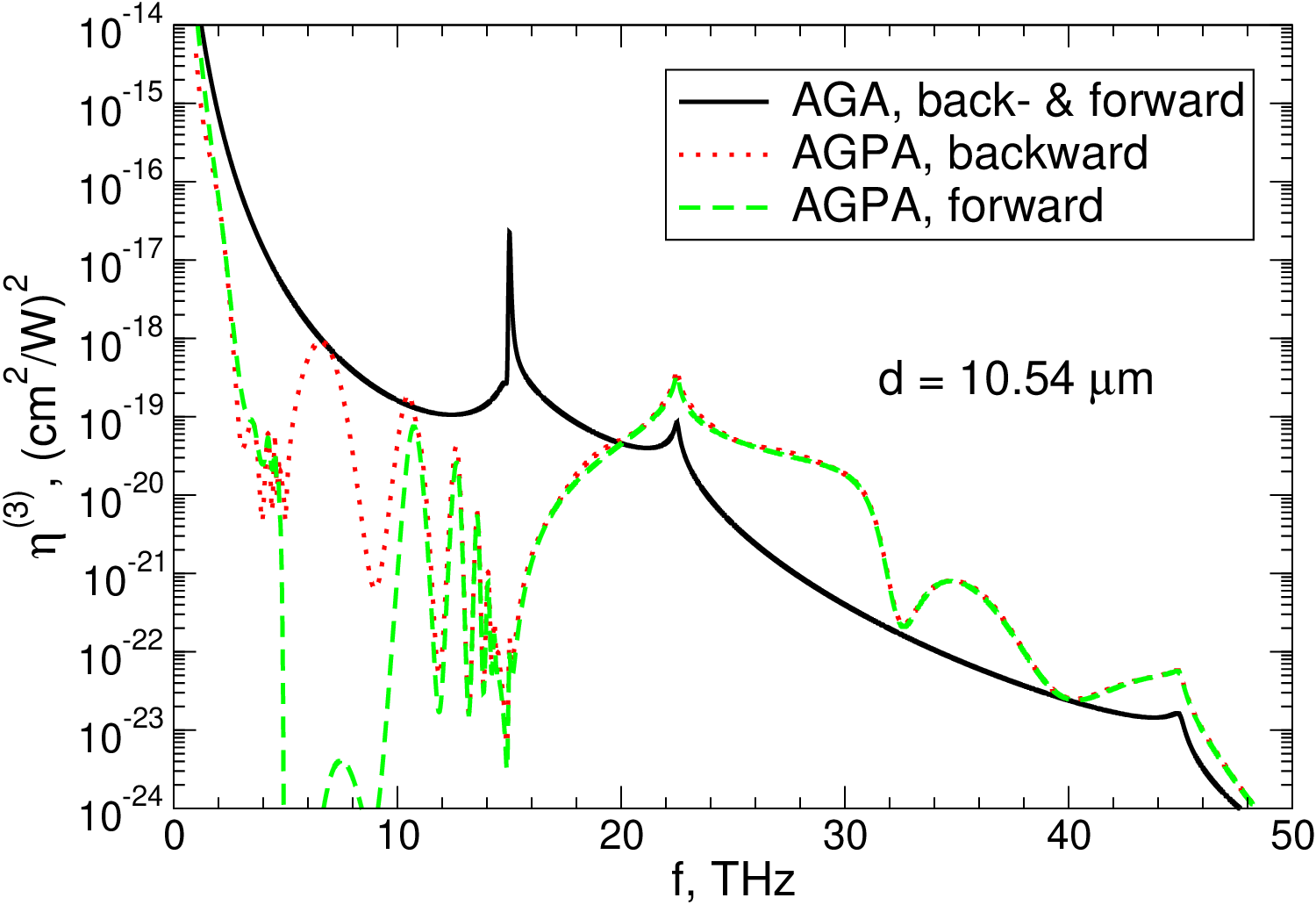}
\caption{\label{fig:AGPA-15THz} The parameter $\eta^{(3)}$ of the AGPA structure as a function of the input-wave frequency $f$ at the polar dielectric thickness $d=10.54$ $\mu$m in the case when the resonance frequency (\ref{Wres}) is close to the TO-phonon frequency $f_{\rm TO}=15$ THz. Parameters of graphene are $\tau=1$ ps and $n_s=0.636\times 10^{12}$ cm$^{-2}$, parameters of the polar dielectric: $\epsilon_\infty=1$, $f_{\rm TO}=15$ THz, $f_{\rm LO}=30$ THz, $\gamma_{\rm TO}/2\pi=0.2$ THz. The black curves show for comparison the result for the isolated graphene layer (AGA). The red dotted and green dashed curves show the third-harmonic wave intensity emitted in the backward and forward direction respectively.}
\end{figure}

Now consider the case when the resonance frequency $\omega_{\rm res}$, Eq. (\ref{Wres}), is close to the TO-phonon frequency. Figure \ref{fig:AGPA-15THz} shows the dependence of the parameter $\eta^{(3)}$ of the AGPA structure on the input-wave frequency $f$ at $d=10.54$ $\mu$m, $\tau=1$ ps and $n_s=0.636\times 10^{12}$ cm$^{-2}$. The density of electrons is chosen so that the resonance frequency (\ref{Wres}) corresponding to the condition $\hbar\omega=2E_F/3$ is close to the TO-phonon frequency $f_{\rm TO}=15$ THz. The general behavior of $\eta^{(3)}$ is similar to that shown in Figure \ref{fig:AGPA-1a}. Near the graphene resonance at 15 THz the third harmonic intensity is strongly (by many orders of magnitude) suppressed, but at the larger frequencies ($\gtrsim 20$ THz) the presence of the polar-dielectric substrate increases the isolated graphene response (by a factor $\simeq 50$ at 30 THz and by a factor of $\simeq 10$ at 35 THz).  

As seen from Figures \ref{fig:AGPA-1a} and \ref{fig:AGPA-15THz}, in the cases considered in Sections \ref{3resTO} and \ref{resTO}, the third harmonic intensity is strongly suppressed by the polar dielectric substrate, if the incident wave frequency is close to the TO-phonon frequency, and substantially enhanced, if it is close to the LO-phonon frequency. Let us now consider what happens if the triple resonance frequency $3\omega_{\rm res}$ is close to the LO-phonon frequency.

\subsubsection{The case $3\omega_{\rm res}\simeq \omega_{\rm LO}$\label{3resLO}}

\begin{figure}
\includegraphics[width=0.495\textwidth]{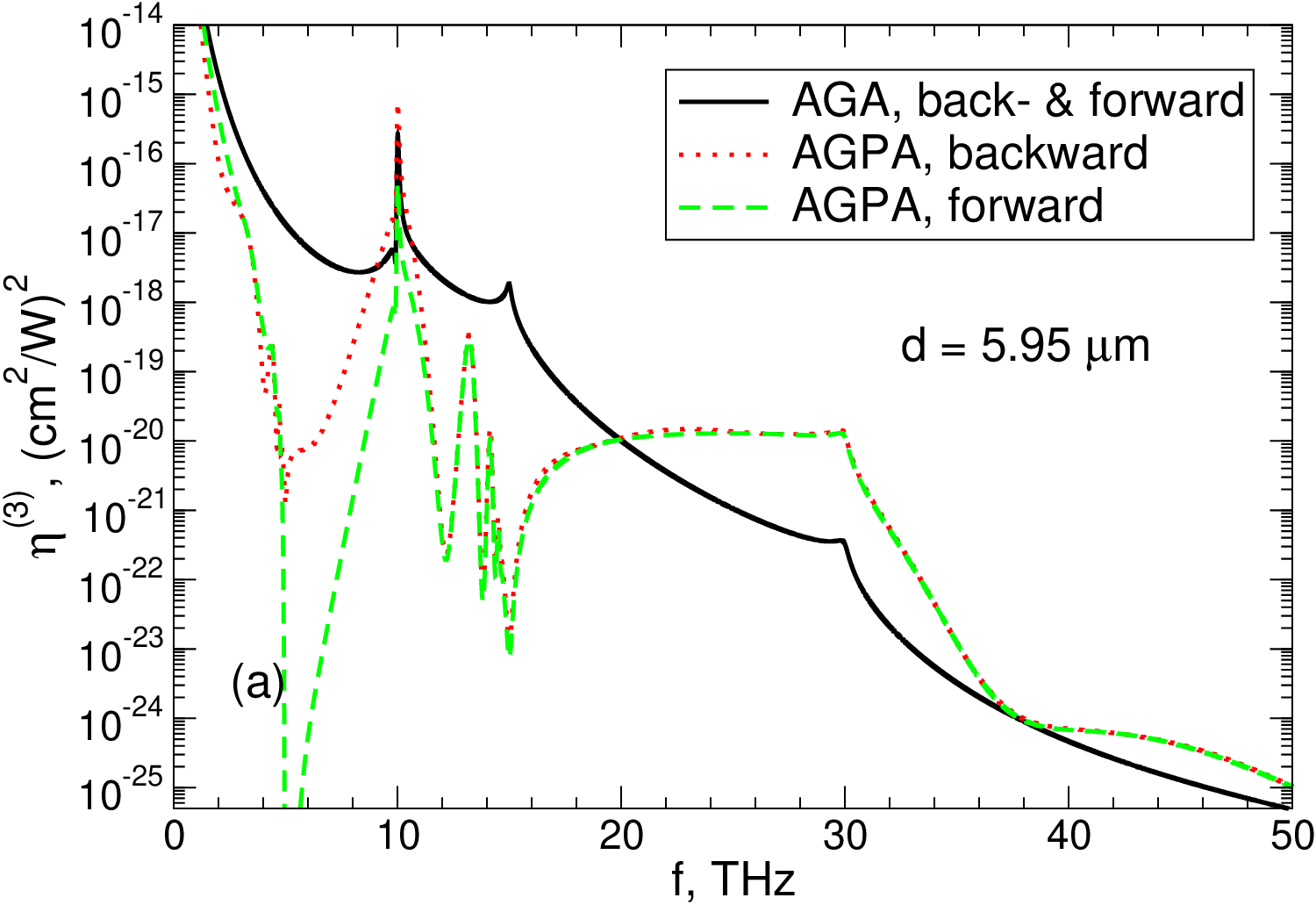}
\includegraphics[width=0.495\textwidth]{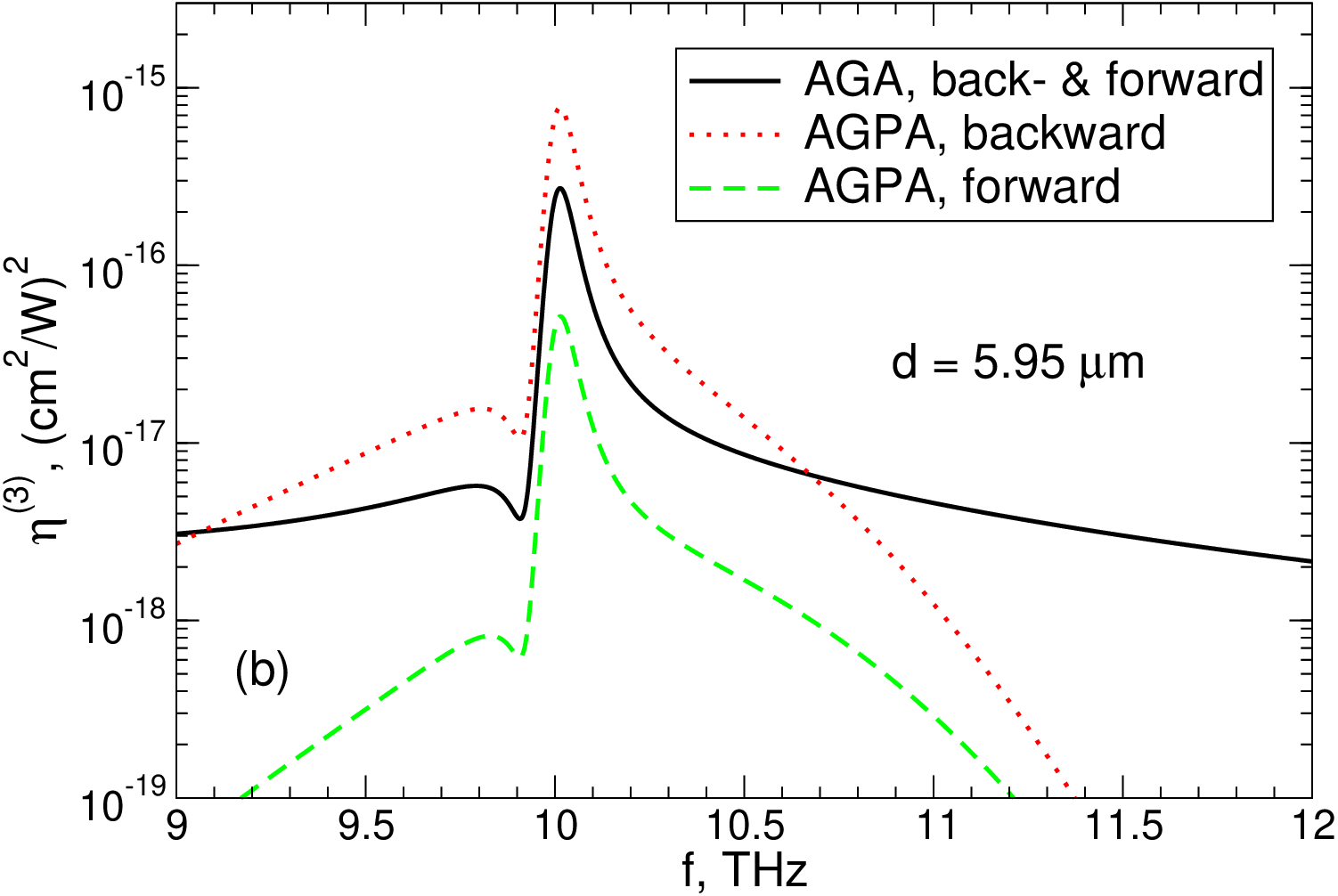}
\includegraphics[width=0.495\textwidth]{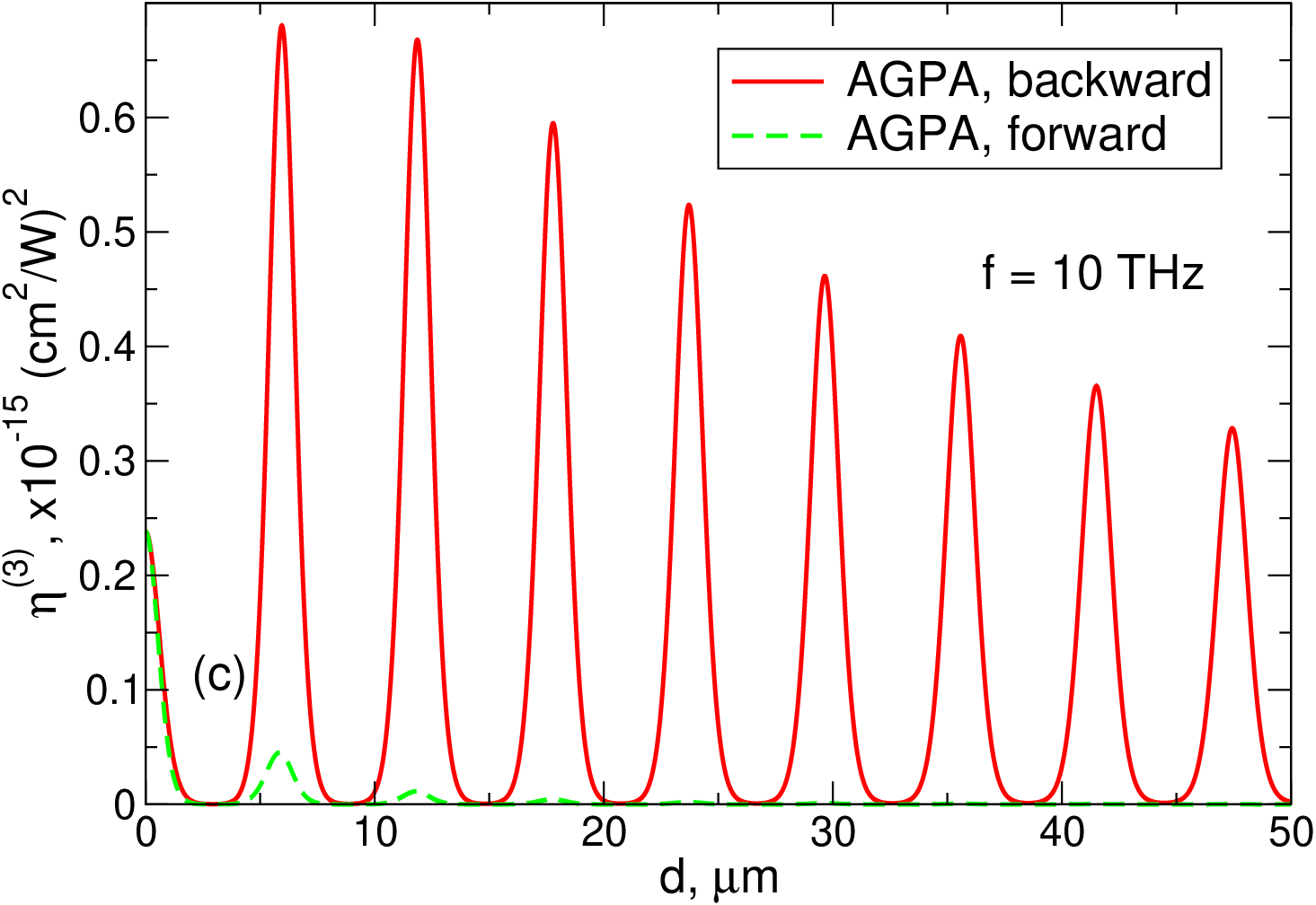}
\caption{\label{fig:AGPA-10THz} (a,b) The parameter $\eta^{(3)}$ of the AGPA structure as a function of the input-wave frequency $f$ at the polar dielectric thickness $d=5.95$ $\mu$m in the case when the triple resonance frequency (\ref{Wres}) is close to the LO-phonon frequency $f_{\rm LO}=30$ THz; (a) -- in a broad frequency range up to 50 THz, (b) in the vicinity of the graphene resonance (\ref{Wres}) at 10 THz. (c) The efficiency $\eta^{(3)}$ as a function of $d$ at $f=10$ THz. Parameters of graphene are $\tau=1$ ps and $n_s=0.2827\times 10^{12}$ cm$^{-2}$, parameters of the polar dielectric: $\epsilon_\infty=1$, $f_{\rm TO}=15$ THz, $f_{\rm LO}=30$ THz, $\gamma_{\rm TO}/2\pi=0.2$ THz. The black curves in (a) and (b) show for comparison the result for the isolated graphene layer (AGA). The red dotted and green dashed curves show the third-harmonic wave intensity emitted in the backward and forward direction respectively.}
\end{figure}

Figure \ref{fig:AGPA-10THz} demonstrates the input-frequency dependence of the parameter $\eta^{(3)}$ in the AGPA structure at the electron density $n_s=0.2827\times 10^{12}$ cm$^{-2}$. The resonance frequency (\ref{Wres}) equals 10 THz in this case, then $3f_{\rm res}=f_{\rm LO}=30$ THz. One sees that at frequencies higher than $\simeq 20$ THz the behavior of $\eta^{(3)}$ is similar to the one observed in two previous cases: the presence on the substrate quite substantially increases the third harmonic intensity and flattens the frequency dependence of $\eta^{(3)}$ in the range $\omega_{\rm TO}\lesssim\omega\lesssim\omega_{\rm LO}$. The strong suppression of $\eta^{(3)}_{\rm AGPA}$ at the frequencies between 5 and 10 THz is due to the fact that the triple frequency lies in the Reststrahlen-Band, and the oscillations of $\eta^{(3)}_{\rm AGPA}$ between 10 and 15 THz have the same origin as the oscillations of the TRA coefficients shown in Figures \ref{fig:AGPA-lin}(b)-(d).

What is new in Figure \ref{fig:AGPA-10THz} as compared to the previous cases is that the third-harmonic intensity (at the frequency $3f\simeq f_{\rm LO}\simeq 30$ THz)  is increased, when the input-wave frequency is close to the graphene resonance (\ref{Wres}), $f\simeq f_{\rm res}\simeq 10$ THz. Figure \ref{fig:AGPA-10THz}(b) shows details of the frequency dependence of $\eta^{(3)}$ around this point. For the backward emitted third harmonic the factor $\eta^{(3)}$ is increased by a factor of $\simeq 2.86$ under the chosen conditions. This is due to the interference of the incident ($\omega$) wave in the dielectric, as seen from Figure \ref{fig:AGPA-10THz}(c) where the dielectric thickness dependence is shown. The oscillation maxima of $\eta^{(3)}$ in this Figure correspond to the interference condition (\ref{interfer-cond1}) where $n_\omega \approx 2.518$ is the refractive index of the substrate at 10 THz. As seen from Figure \ref{fig:AGPA-10THz}(c) the substrates of \textit{certain} thicknesses increase the effect, however, if the optimal conditions are not satisfied, the effect can be strongly suppressed as compared to the case $d=0$.

\subsubsection{The case $\omega_{\rm res}\simeq \omega_{\rm LO}$}

\begin{figure}
\includegraphics[width=0.495\textwidth]{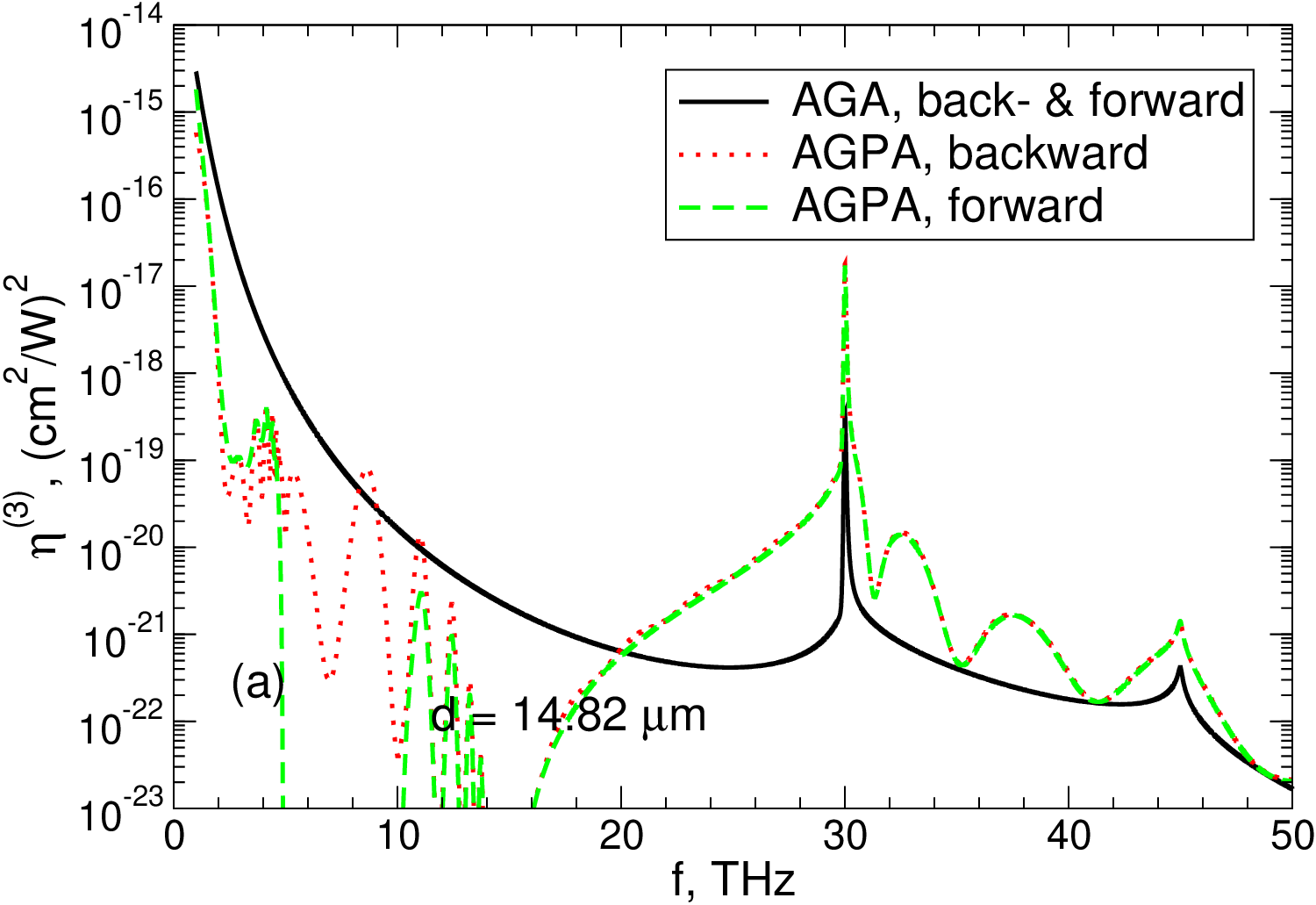}
\includegraphics[width=0.495\textwidth]{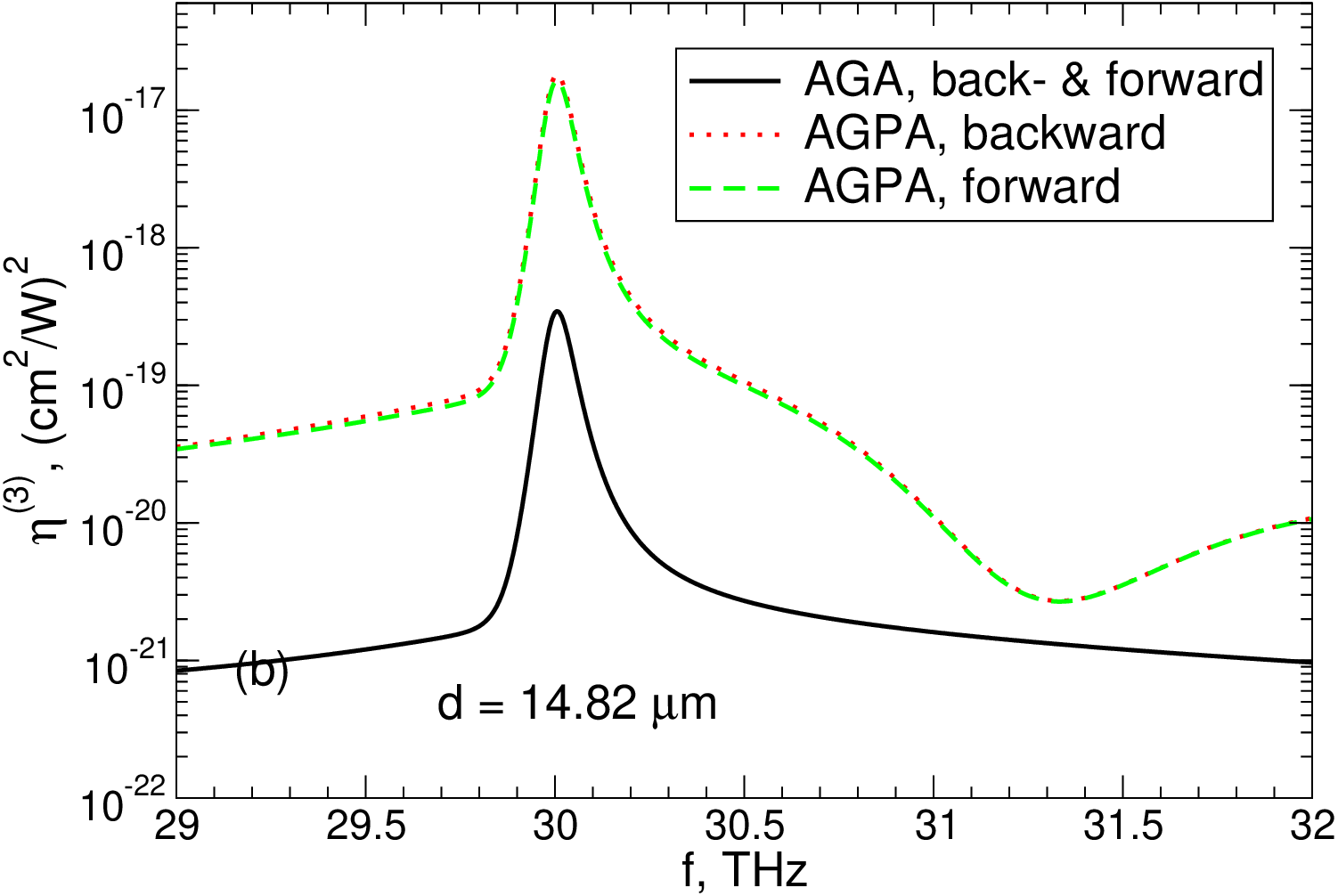}
\includegraphics[width=0.495\textwidth]{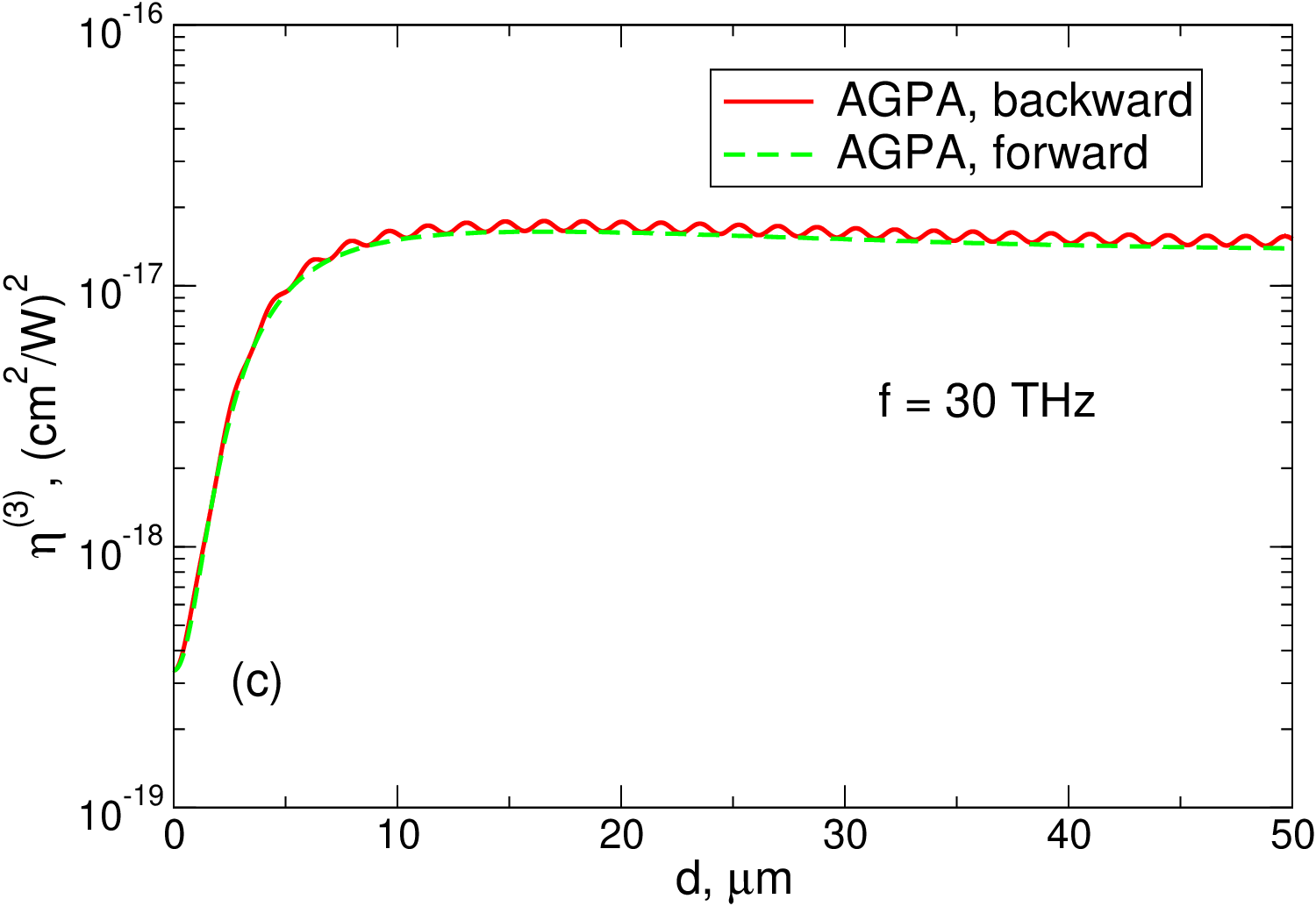}
\includegraphics[width=0.495\textwidth]{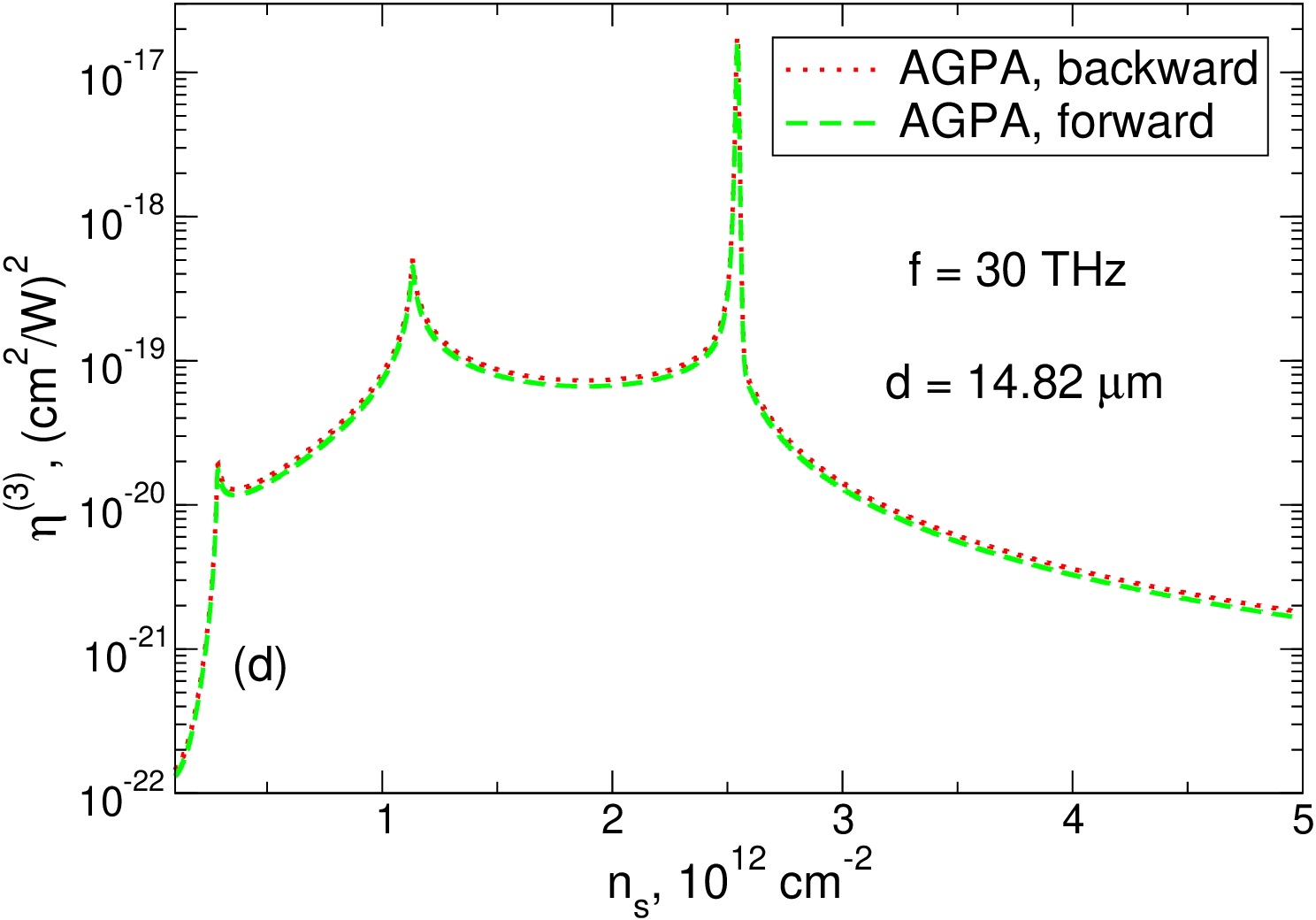}
\caption{\label{fig:AGPA-30THz} (a,b) The parameter $\eta^{(3)}$ of the AGPA structure as a function of the input-wave frequency $f$ at the polar dielectric thickness $d=14.82$ $\mu$m in the case when the graphene resonance frequency (\ref{Wres}) is close to the LO-phonon frequency of the substrate $f_{\rm LO}=30$ THz; (a) -- in a broad frequency range up to 50 THz, (b) in the vicinity of the graphene resonance at 30 THz. (c) The same parameter as a function of the dielectric thickness at $f=30$ THz. Parameters of graphene in Figures (a)--(c) are $\tau=1$ ps and $n_s=2.54469\times 10^{12}$ cm$^{-2}$. (d) The efficiency $\eta^{(3)}$ as a function of the electron density at $d=14.82$ $\mu$m and $f=30$ THz. Parameters of the polar dielectric in all panels: $\epsilon_\infty=1$, $f_{\rm TO}=15$ THz, $f_{\rm LO}=30$ THz, $\gamma_{\rm TO}/2\pi=0.2$ THz. The black curves in (a) and (b) show for comparison the result for the isolated graphene layer (AGA). The red dotted and green dashed curves show the third-harmonic wave intensity emitted in the backward and forward direction respectively.}
\end{figure}

Finally we consider the case when the graphene resonance (\ref{Wres}) coincides with the LO-phonon resonance, Figure \ref{fig:AGPA-30THz}. In this case the density of electrons is about $2.5447\times 10^{12}$ cm$^{-2}$ which gives the resonance position at $\simeq 30$ THz. As seen from Figure \ref{fig:AGPA-30THz}(a) and especially Figure \ref{fig:AGPA-30THz}(b), the combination of two resonances helps to substantially increase the third-harmonic (90 THz) intensity if the input-wave frequency is close to the LO-phonon frequency and to the graphene resonance frequency $f_{\rm res}$ (30 THz in our example). The amplification factor, under the chosen conditions, is about 52. The $d$-dependence of the up-conversion factor $\eta^{(3)}$ at $f=30$ THz is shown in Figure \ref{fig:AGPA-30THz}(c). One sees that $\eta^{(3)}$ monotonously grows with $d$ when $d\lesssim 10$ $\mu$m and then practically saturates. The polar dielectric substrate thus strongly increases the effect if $d\lesssim 10$ $\mu$m; at larger $d$ the choice of the dielectric thickness is not very crucial. The intensity of the backward emitted third harmonic slightly oscillates with $d$ while that of the forward emitted wave does not (see the discussion of Figure \ref{fig:AGPA-1b}(b)). Notice that at the parameters of Figure \ref{fig:AGPA-30THz}(a) one of the interference maxima also coincides with the second graphene resonance $\hbar\omega=E_F$ at $f\simeq 45$ THz. The third-harmonic response is also resonantly enhanced near this frequency, from $\eta^{(3)}_{\rm AGA}\simeq 4.19\times 10^{-22}$ up to $\eta^{(3)}_{\rm AGPA}\simeq 1.43\times 10^{-21}$ (cm$^2$/W)$^2$ (the enhancement factor is $\simeq 3.4$).

If the graphene resonance frequency (\ref{Wres}) slightly deviates from $\omega_{\rm LO}$ the intensity of the third harmonic falls down drastically. This is illustrated in Figure \ref{fig:AGPA-30THz}(d) where the dependence of $\eta^{(3)}_{\rm AGPA}$ on the electron density, and hence on the resonance frequency $f_{\rm res}$, is shown at the input-wave frequency $f=30$ THz (which is equal to $f_{\rm LO}$ in this case) and the polar dielectric thickness $d=14.82$ $\mu$m. One sees that the resonance around the density $n_s= 2.54469\times 10^{12}$ cm$^{-2}$ is very sharp; small deviations from the resonance number leads to a strong suppression of the effect. Two other resonances at $n_s=1.13\times 10^{12}$ cm$^{-2}$ and $n_s=0.28\times 10^{12}$ cm$^{-2}$ correspond to the graphene resonances $\hbar\omega=E_F$ and $\hbar\omega=2E_F$ respectively. The sharpness of the resonances also depends on the relaxation time $\tau$; if it is smaller than 1 ps, the height of the resonances will be smaller, compare with Figure \ref{fig:AGA}(a).

\subsection{Structure AGPMA \label{sec:agpma}}

\begin{figure}
\includegraphics[width=0.495\textwidth]{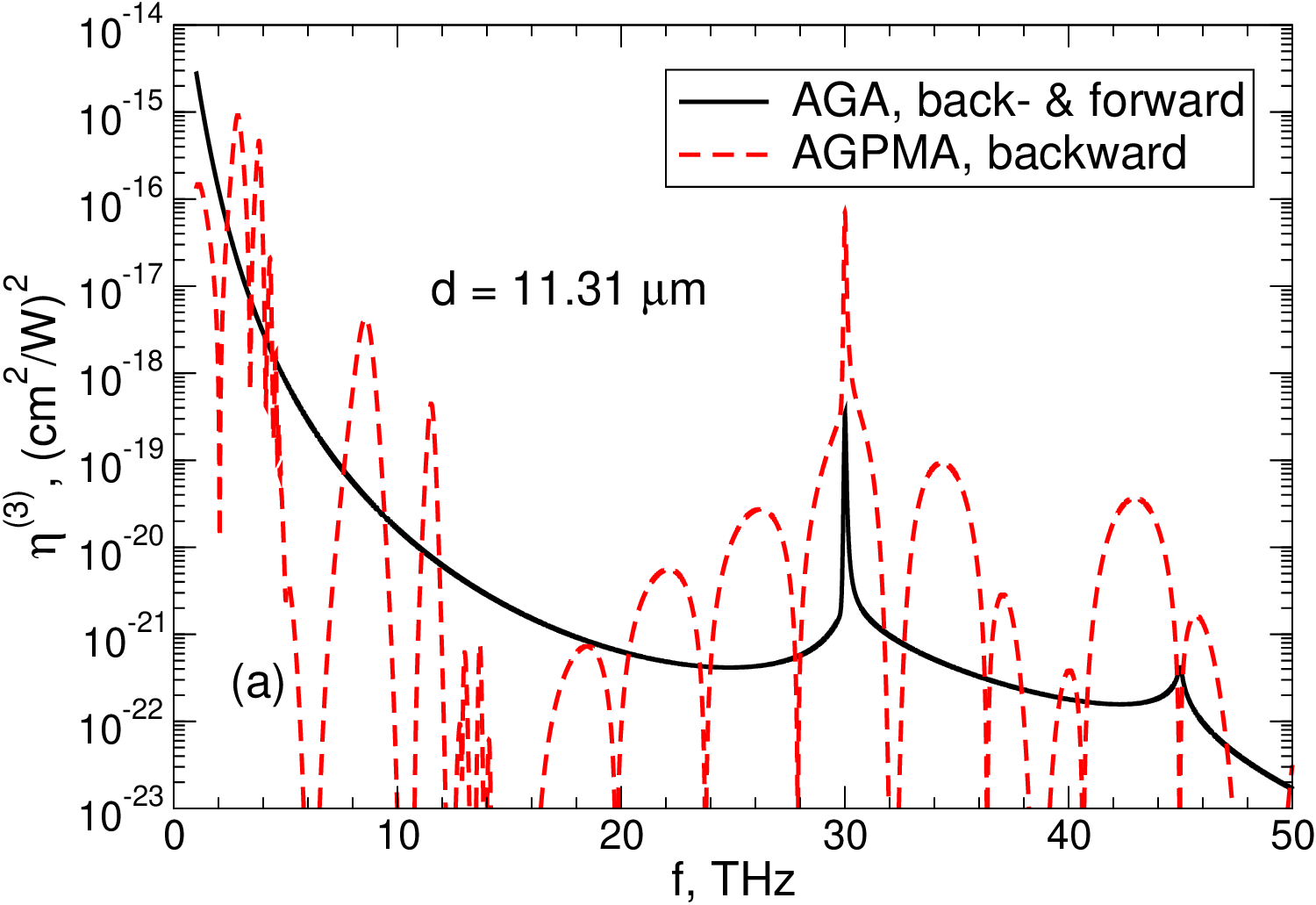}
\includegraphics[width=0.495\textwidth]{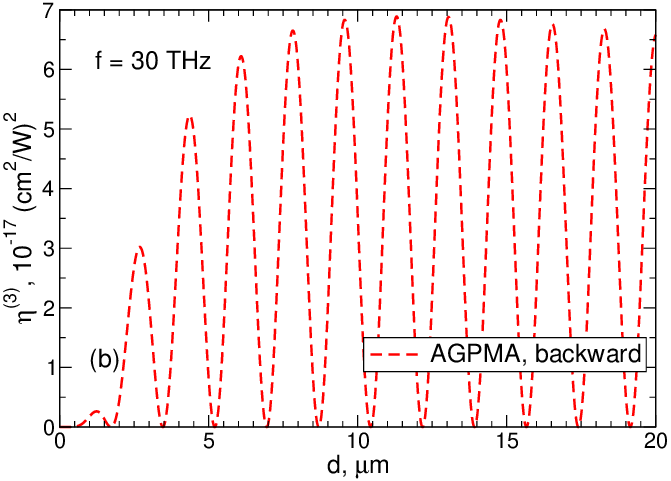}
\caption{\label{fig:AGPM-30THz} The parameter $\eta^{(3)}$ of the AGPMA structure (a) as a function of the input-wave frequency $f$ at the polar dielectric thickness $d=11.31$ $\mu$m in the case when the graphene resonance frequency (\ref{Wres}) is close to the LO-phonon frequency of the substrate $f_{\rm LO}=30$ THz and (b) as a function of $d$ at $f=30$ THz. Parameters of graphene are $\tau=1$ ps and $n_s=2.54469\times 10^{12}$ cm$^{-2}$, parameters of the polar dielectric: $\epsilon_\infty=1$, $f_{\rm TO}=15$ THz, $f_{\rm LO}=30$ THz, $\gamma_{\rm TO}/2\pi=0.2$ THz, the metal thickness is $0.2$ $\mu$m. The black curve in (a) shows for comparison the result for the isolated graphene layer (AGA). The red dashed curve shows the third-harmonic wave intensity emitted in the backward direction.}
\end{figure}

Now we study how the thin metallic layer on the backside of the polar dielectric (the structure AGPMA) may influence the up-conversion efficiency $\eta^{(3)}$ as compared to the AGA structure. Figure \ref{fig:AGPM-30THz}(a) shows the frequency dependence of $\eta^{(3)}$ under the condition $\omega_{\rm res}=\omega_{\rm LO}$, which corresponds to the electron density $n_s=2.54469\times 10^{12}$ cm$^{-2}$. The polar dielectric thickness is chosen to be $d=11.31$ $\mu$m which corresponds to the highest interferometric resonance at $30$ THz, see Figure \ref{fig:AGPM-30THz}(b). One sees that, due to many intereferometric maxima, the spectrum of $\eta^{(3)}_{\rm AGPMA}$ has many frequency bands in which $\eta^{(3)}_{\rm AGPMA}>\eta^{(3)}_{\rm AGA}$. At $d=11.31$ $\mu$m one of these maxima coincides with the graphene resonance (\ref{Wres}) at $f=30$ THz. The value of $\eta^{(3)}_{\rm AGPMA}$ in this point, $\simeq 7.18\times 10^{-17}$ (cm$^2$/W)$^2$, is more than 200 times larger than $\eta^{(3)}_{\rm AGA}\approx 3.47\times 10^{-19}$ (cm$^2$/W)$^2$. Like in the case of the AGDMA structures, the metalization of the back-side of the polar dielectric also leads to a strong enhancement, under certain conditions, of the third-harmonic generation from the AGPMA structure. The correct choice of the dielectric thickness is thus crucial for a successful experiment. Notice that at small $d$ ($\lesssim 1$ $\mu$m) the effect practically disappears due to the screening of the electric fields by metal.

\subsection{Terahertz response of a AGDMA structure}

Finally we show results obtained for the low-frequency (one to a few THz) response of the AGDMA structures. At such low frequencies the third-order response of graphene is not resonant, since at realistic electron densities the Fermi energy is typically larger than $\hbar\omega$. On the other hand, the absolute value of $\eta^{(3)}$ grows when the frequency decreases, therefore it makes sense to quantitatively investigate the role of the dielectric substrate at $f\simeq 1$ THz. At such low frequencies the dielectric permittivity of the polar dielectric can be considered to be frequency independent, therefore we ignore the difference between the AGPMA and AGDMA structures. 

\begin{figure}
\includegraphics[width=0.495\textwidth]{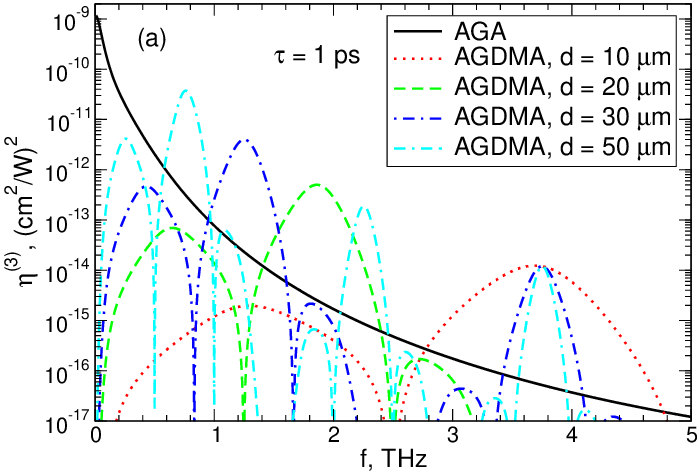}
\includegraphics[width=0.495\textwidth]{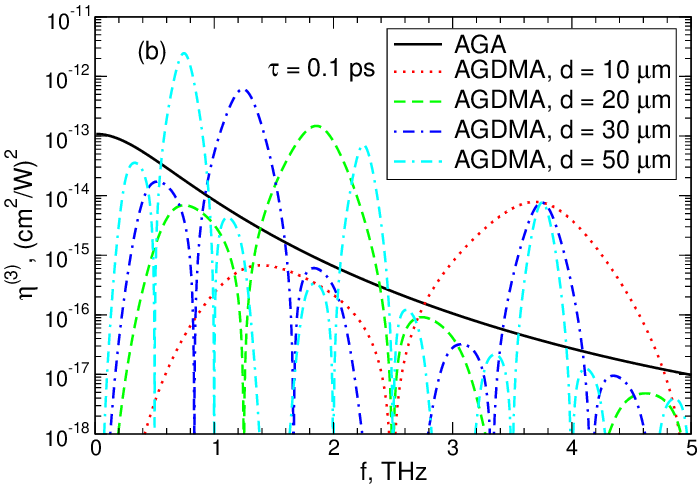}
\caption{\label{fig:AGDM-THz} The parameter $\eta^{(3)}$ of the AGDMA structure as a function of the input-wave frequency $f$ at different values of the dielectric thickness $d$ and at the relaxation times (a) $\tau=1$ ps and (b) $\tau=0.1$ ps. The density of electrons in graphene is $n_s=0.3\times 10^{12}$ cm$^{-2}$, the refractive index of the dielectric substrate is $n_\omega=n_{3\omega}=2$, the metal (Au) thickness is $0.4$ $\mu$m. The black curves show for comparison the result for the isolated graphene layer (AGA). Only the emission in the backward direction is shown for all curves (the emission in the forward direction is negligibly small).}
\end{figure}

Figure \ref{fig:AGDM-THz} shows the frequency dependence of the up-conversion parameter $\eta^{(3)}_{\rm AGDMA}$ at different values of the dielectric thickness and two values of the graphene relaxation time, $\tau=1$ ps (Figure \ref{fig:AGDM-THz}(a)) and $\tau=0.1$ ps (Figure \ref{fig:AGDM-THz}(b)). The back side of the dielectric is assumed to be metalized by gold of the thickness $0.4$ $\mu$m, the refractive index $n_\omega=n_{3\omega}=2$ is considered to be frequency independent. One sees that the third-harmonic intensity can be more than two orders of magnitude larger than in the AGA structure and this enhancement can be achieved in different frequency intervals by choosing different substrate thicknesses. The absolute values of the parameter $\eta^{(3)}$ at the input-wave frequency $f\simeq 1$ THz can be as large as $\sim 5\times 10^{-11}$ (cm$^2$/W)$^2$ in structures with the graphene relaxation time $\tau=1$ ps and $\sim 3\times 10^{-12}$ (cm$^2$/W)$^2$ at $\tau=0.1$ ps. This corresponds to the output signal intensity (at $3f\simeq 3$ THz) of 50 and 3 mW/cm$^2$ respectively, at the input-wave intensity of only 1 kW/cm$^2$.

\section{Summary and conclusions\label{sec:summ}}

We have investigated the third harmonic generation effect from structures consisting of the nonlinear material graphene lying on different substrates. We have shown that, dependent on the type and thickness of the substrate and/or on a frequency interval, the third harmonic intensity can be both drastically suppressed and substantially enhanced as compared to the freely hanging graphene. The most essential growth of the third harmonic is shown to exist in the structures with the metalized back side and in structures with the polar dielectric serving as a substrate. In the first case the strong enhancement of the third harmonic is caused by the interference of the input ($\omega$) and output ($3\omega$) waves in the dielectric substrate, due to the reflection from the back-side metallic mirror. In the second case the growth of the third harmonic is caused by the interaction of either the input ($\omega$) and output ($3\omega$) waves with the LO phonons in the polar dielectric substrate. A correct choice of the substrate thickness is thus crucially important for the optimization of the emitted third-harmonic intensity.

Another interesting effect that we have found is a flattening, in certain frequency intervals, of the frequency dependence of the graphene-on-a-substrate system, as compared to the very strong frequency dependence of the pure-graphene response. We have shown that, if graphene lies on the polar-dielectric substrate, this strong frequency dependence can be compensated by the dispersion of the dielectric permittivity of the substrate, so that in the quite broad frequency interval between the TO and LO phonon frequencies, $\omega_{\rm TO}\lesssim \omega\lesssim \omega_{\rm LO}$, the resulting intensity of the third harmonic, at $3\omega_{\rm TO}\lesssim 3\omega\lesssim 3\omega_{\rm LO}$, turns out almost frequency independent. 

To summarize, our work shows that the strongly nonlinear electrodynamic properties of graphene can be further increased by a proper design of the underlying substrate. Results of this work can be used in different terahertz and optoelectronic applications.

\acknowledgments

The work has received funding from the European Union's Horizon 2020 research and innovation programme GrapheneCore1 under Grant Agreement No. 696656.


\begin{thebibliography}{45}
\expandafter\ifx\csname natexlab\endcsname\relax\def\natexlab#1{#1}\fi
\expandafter\ifx\csname bibnamefont\endcsname\relax
  \def\bibnamefont#1{#1}\fi
\expandafter\ifx\csname bibfnamefont\endcsname\relax
  \def\bibfnamefont#1{#1}\fi
\expandafter\ifx\csname citenamefont\endcsname\relax
  \def\citenamefont#1{#1}\fi
\expandafter\ifx\csname url\endcsname\relax
  \def\url#1{\texttt{#1}}\fi
\expandafter\ifx\csname urlprefix\endcsname\relax\def\urlprefix{URL }\fi
\providecommand{\bibinfo}[2]{#2}
\providecommand{\eprint}[2][]{\url{#2}}

\bibitem[{\citenamefont{Mikhailov}(2007)}]{Mikhailov07e}
\bibinfo{author}{\bibfnamefont{S.~A.} \bibnamefont{Mikhailov}},
  \bibinfo{journal}{Europhys. Lett.} \textbf{\bibinfo{volume}{79}},
  \bibinfo{pages}{27002} (\bibinfo{year}{2007}).

\bibitem[{\citenamefont{Dragoman et~al.}(2010)\citenamefont{Dragoman, Neculoiu,
  Deligeorgis, Konstantinidis, Dragoman, Cismaru, Muller, and
  Plana}}]{Dragoman10}
\bibinfo{author}{\bibfnamefont{M.}~\bibnamefont{Dragoman}},
  \bibinfo{author}{\bibfnamefont{D.}~\bibnamefont{Neculoiu}},
  \bibinfo{author}{\bibfnamefont{G.}~\bibnamefont{Deligeorgis}},
  \bibinfo{author}{\bibfnamefont{G.}~\bibnamefont{Konstantinidis}},
  \bibinfo{author}{\bibfnamefont{D.}~\bibnamefont{Dragoman}},
  \bibinfo{author}{\bibfnamefont{A.}~\bibnamefont{Cismaru}},
  \bibinfo{author}{\bibfnamefont{A.~A.} \bibnamefont{Muller}},
  \bibnamefont{and} \bibinfo{author}{\bibfnamefont{R.}~\bibnamefont{Plana}},
  \bibinfo{journal}{Appl. Phys. Lett.} \textbf{\bibinfo{volume}{97}},
  \bibinfo{pages}{093101} (\bibinfo{year}{2010}).

\bibitem[{\citenamefont{Hendry et~al.}(2010)\citenamefont{Hendry, Hale, Moger,
  Savchenko, and Mikhailov}}]{Hendry10}
\bibinfo{author}{\bibfnamefont{E.}~\bibnamefont{Hendry}},
  \bibinfo{author}{\bibfnamefont{P.~J.} \bibnamefont{Hale}},
  \bibinfo{author}{\bibfnamefont{J.~J.} \bibnamefont{Moger}},
  \bibinfo{author}{\bibfnamefont{A.~K.} \bibnamefont{Savchenko}},
  \bibnamefont{and} \bibinfo{author}{\bibfnamefont{S.~A.}
  \bibnamefont{Mikhailov}}, \bibinfo{journal}{Phys. Rev. Lett.}
  \textbf{\bibinfo{volume}{105}}, \bibinfo{pages}{097401}
  (\bibinfo{year}{2010}).

\bibitem[{\citenamefont{Bykov et~al.}(2012)\citenamefont{Bykov, Murzina, Rybin,
  and Obraztsova}}]{Bykov12}
\bibinfo{author}{\bibfnamefont{A.~Y.} \bibnamefont{Bykov}},
  \bibinfo{author}{\bibfnamefont{T.~V.} \bibnamefont{Murzina}},
  \bibinfo{author}{\bibfnamefont{M.~G.} \bibnamefont{Rybin}}, \bibnamefont{and}
  \bibinfo{author}{\bibfnamefont{E.~D.} \bibnamefont{Obraztsova}},
  \bibinfo{journal}{Phys. Rev. B} \textbf{\bibinfo{volume}{85}},
  \bibinfo{pages}{121413(R)} (\bibinfo{year}{2012}).

\bibitem[{\citenamefont{Kumar et~al.}(2013)\citenamefont{Kumar, Kumar,
  Gerstenkorn, Wang, Chiu, Smirl, and Zhao}}]{Kumar13}
\bibinfo{author}{\bibfnamefont{N.}~\bibnamefont{Kumar}},
  \bibinfo{author}{\bibfnamefont{J.}~\bibnamefont{Kumar}},
  \bibinfo{author}{\bibfnamefont{C.}~\bibnamefont{Gerstenkorn}},
  \bibinfo{author}{\bibfnamefont{R.}~\bibnamefont{Wang}},
  \bibinfo{author}{\bibfnamefont{H.-Y.} \bibnamefont{Chiu}},
  \bibinfo{author}{\bibfnamefont{A.~L.} \bibnamefont{Smirl}}, \bibnamefont{and}
  \bibinfo{author}{\bibfnamefont{H.}~\bibnamefont{Zhao}},
  \bibinfo{journal}{Phys. Rev. B} \textbf{\bibinfo{volume}{87}},
  \bibinfo{pages}{121406(R)} (\bibinfo{year}{2013}).

\bibitem[{\citenamefont{Hong et~al.}(2013)\citenamefont{Hong, Dadap, Petrone,
  Yeh, Hone, and {Osgood, Jr.}}}]{Hong13}
\bibinfo{author}{\bibfnamefont{S.-Y.} \bibnamefont{Hong}},
  \bibinfo{author}{\bibfnamefont{J.~I.} \bibnamefont{Dadap}},
  \bibinfo{author}{\bibfnamefont{N.}~\bibnamefont{Petrone}},
  \bibinfo{author}{\bibfnamefont{P.-C.} \bibnamefont{Yeh}},
  \bibinfo{author}{\bibfnamefont{J.}~\bibnamefont{Hone}}, \bibnamefont{and}
  \bibinfo{author}{\bibfnamefont{R.~M.} \bibnamefont{{Osgood, Jr.}}},
  \bibinfo{journal}{Phys. Rev. X} \textbf{\bibinfo{volume}{3}},
  \bibinfo{pages}{021014} (\bibinfo{year}{2013}).

\bibitem[{\citenamefont{Hotopan et~al.}(2011)\citenamefont{Hotopan, {Ver
  Hoeye}, Vazquez, Camblor, Fern\'andez, Las~Heras, \'Alvarez, and
  Men\'endez}}]{Hotopan11}
\bibinfo{author}{\bibfnamefont{G.}~\bibnamefont{Hotopan}},
  \bibinfo{author}{\bibfnamefont{S.}~\bibnamefont{{Ver Hoeye}}},
  \bibinfo{author}{\bibfnamefont{C.}~\bibnamefont{Vazquez}},
  \bibinfo{author}{\bibfnamefont{R.}~\bibnamefont{Camblor}},
  \bibinfo{author}{\bibfnamefont{M.}~\bibnamefont{Fern\'andez}},
  \bibinfo{author}{\bibfnamefont{F.}~\bibnamefont{Las~Heras}},
  \bibinfo{author}{\bibfnamefont{P.}~\bibnamefont{\'Alvarez}},
  \bibnamefont{and}
  \bibinfo{author}{\bibfnamefont{R.}~\bibnamefont{Men\'endez}},
  \bibinfo{journal}{Progress In Electromagnetic Research}
  \textbf{\bibinfo{volume}{118}}, \bibinfo{pages}{57} (\bibinfo{year}{2011}).

\bibitem[{\citenamefont{Gu et~al.}(2012)\citenamefont{Gu, Petrone, McMillan,
  {van der Zande}, Yu, Lo, Kwong, Hone, and Wong}}]{Gu12}
\bibinfo{author}{\bibfnamefont{T.}~\bibnamefont{Gu}},
  \bibinfo{author}{\bibfnamefont{N.}~\bibnamefont{Petrone}},
  \bibinfo{author}{\bibfnamefont{J.~F.} \bibnamefont{McMillan}},
  \bibinfo{author}{\bibfnamefont{A.}~\bibnamefont{{van der Zande}}},
  \bibinfo{author}{\bibfnamefont{M.}~\bibnamefont{Yu}},
  \bibinfo{author}{\bibfnamefont{G.~Q.} \bibnamefont{Lo}},
  \bibinfo{author}{\bibfnamefont{D.~L.} \bibnamefont{Kwong}},
  \bibinfo{author}{\bibfnamefont{J.}~\bibnamefont{Hone}}, \bibnamefont{and}
  \bibinfo{author}{\bibfnamefont{C.~W.} \bibnamefont{Wong}},
  \bibinfo{journal}{Nature Photonics} \textbf{\bibinfo{volume}{6}},
  \bibinfo{pages}{554} (\bibinfo{year}{2012}).

\bibitem[{\citenamefont{Zhang et~al.}(2012)\citenamefont{Zhang, Virally, Bao,
  Ping, Massar, Godbout, and Kockaert}}]{Zhang12}
\bibinfo{author}{\bibfnamefont{H.}~\bibnamefont{Zhang}},
  \bibinfo{author}{\bibfnamefont{S.}~\bibnamefont{Virally}},
  \bibinfo{author}{\bibfnamefont{Q.}~\bibnamefont{Bao}},
  \bibinfo{author}{\bibfnamefont{L.~K.} \bibnamefont{Ping}},
  \bibinfo{author}{\bibfnamefont{S.}~\bibnamefont{Massar}},
  \bibinfo{author}{\bibfnamefont{N.}~\bibnamefont{Godbout}}, \bibnamefont{and}
  \bibinfo{author}{\bibfnamefont{P.}~\bibnamefont{Kockaert}},
  \bibinfo{journal}{Optics Letters} \textbf{\bibinfo{volume}{37}},
  \bibinfo{pages}{1856} (\bibinfo{year}{2012}).

\bibitem[{\citenamefont{Popa et~al.}(2010)\citenamefont{Popa, Sun, Torrisi,
  Hasan, Wang, and Ferrari}}]{Popa10}
\bibinfo{author}{\bibfnamefont{D.}~\bibnamefont{Popa}},
  \bibinfo{author}{\bibfnamefont{Z.}~\bibnamefont{Sun}},
  \bibinfo{author}{\bibfnamefont{F.}~\bibnamefont{Torrisi}},
  \bibinfo{author}{\bibfnamefont{T.}~\bibnamefont{Hasan}},
  \bibinfo{author}{\bibfnamefont{F.}~\bibnamefont{Wang}}, \bibnamefont{and}
  \bibinfo{author}{\bibfnamefont{A.~C.} \bibnamefont{Ferrari}},
  \bibinfo{journal}{Appl. Phys. Lett.} \textbf{\bibinfo{volume}{97}},
  \bibinfo{pages}{203106} (\bibinfo{year}{2010}).

\bibitem[{\citenamefont{Popa et~al.}(2011)\citenamefont{Popa, Sun, Hasan,
  Torrisi, Wang, and Ferrari}}]{Popa11}
\bibinfo{author}{\bibfnamefont{D.}~\bibnamefont{Popa}},
  \bibinfo{author}{\bibfnamefont{Z.}~\bibnamefont{Sun}},
  \bibinfo{author}{\bibfnamefont{T.}~\bibnamefont{Hasan}},
  \bibinfo{author}{\bibfnamefont{F.}~\bibnamefont{Torrisi}},
  \bibinfo{author}{\bibfnamefont{F.}~\bibnamefont{Wang}}, \bibnamefont{and}
  \bibinfo{author}{\bibfnamefont{A.~C.} \bibnamefont{Ferrari}},
  \bibinfo{journal}{Appl. Phys. Lett.} \textbf{\bibinfo{volume}{98}},
  \bibinfo{pages}{073106} (\bibinfo{year}{2011}).

\bibitem[{\citenamefont{Mikhailov and Ziegler}(2008)}]{Mikhailov08a}
\bibinfo{author}{\bibfnamefont{S.~A.} \bibnamefont{Mikhailov}}
  \bibnamefont{and} \bibinfo{author}{\bibfnamefont{K.}~\bibnamefont{Ziegler}},
  \bibinfo{journal}{J. Phys. Condens. Matter} \textbf{\bibinfo{volume}{20}},
  \bibinfo{pages}{384204} (\bibinfo{year}{2008}).

\bibitem[{\citenamefont{Mikhailov}(2009{\natexlab{a}})}]{Mikhailov09a}
\bibinfo{author}{\bibfnamefont{S.~A.} \bibnamefont{Mikhailov}},
  \bibinfo{journal}{Microelectron. J.} \textbf{\bibinfo{volume}{40}},
  \bibinfo{pages}{712} (\bibinfo{year}{2009}{\natexlab{a}}).

\bibitem[{\citenamefont{Mikhailov}(2009{\natexlab{b}})}]{Mikhailov09b}
\bibinfo{author}{\bibfnamefont{S.~A.} \bibnamefont{Mikhailov}},
  \bibinfo{journal}{Phys. Rev. B} \textbf{\bibinfo{volume}{79}},
  \bibinfo{pages}{241309(R)} (\bibinfo{year}{2009}{\natexlab{b}}).

\bibitem[{\citenamefont{Dean and van Driel}(2009)}]{Dean09}
\bibinfo{author}{\bibfnamefont{J.~J.} \bibnamefont{Dean}} \bibnamefont{and}
  \bibinfo{author}{\bibfnamefont{H.~M.} \bibnamefont{van Driel}},
  \bibinfo{journal}{Appl. Phys. Lett.} \textbf{\bibinfo{volume}{95}},
  \bibinfo{pages}{261910} (\bibinfo{year}{2009}).

\bibitem[{\citenamefont{Dean and van Driel}(2010)}]{Dean10}
\bibinfo{author}{\bibfnamefont{J.~J.} \bibnamefont{Dean}} \bibnamefont{and}
  \bibinfo{author}{\bibfnamefont{H.~M.} \bibnamefont{van Driel}},
  \bibinfo{journal}{Phys. Rev. B} \textbf{\bibinfo{volume}{82}},
  \bibinfo{pages}{125411} (\bibinfo{year}{2010}).

\bibitem[{\citenamefont{Ishikawa}(2010)}]{Ishikawa10}
\bibinfo{author}{\bibfnamefont{K.~L.} \bibnamefont{Ishikawa}},
  \bibinfo{journal}{Phys. Rev. B} \textbf{\bibinfo{volume}{82}},
  \bibinfo{pages}{201402} (\bibinfo{year}{2010}).

\bibitem[{\citenamefont{Mikhailov}(2011)}]{Mikhailov11c}
\bibinfo{author}{\bibfnamefont{S.~A.} \bibnamefont{Mikhailov}},
  \bibinfo{journal}{Phys. Rev. B} \textbf{\bibinfo{volume}{84}},
  \bibinfo{pages}{045432} (\bibinfo{year}{2011}).

\bibitem[{\citenamefont{Jafari}(2012)}]{Jafari12}
\bibinfo{author}{\bibfnamefont{S.~A.} \bibnamefont{Jafari}},
  \bibinfo{journal}{J. Phys. Condens. Matter} \textbf{\bibinfo{volume}{24}},
  \bibinfo{pages}{205802} (\bibinfo{year}{2012}).

\bibitem[{\citenamefont{Mikhailov and Beba}(2012)}]{Mikhailov12c}
\bibinfo{author}{\bibfnamefont{S.~A.} \bibnamefont{Mikhailov}}
  \bibnamefont{and} \bibinfo{author}{\bibfnamefont{D.}~\bibnamefont{Beba}},
  \bibinfo{journal}{New J. Phys.} \textbf{\bibinfo{volume}{14}},
  \bibinfo{pages}{115024} (\bibinfo{year}{2012}).

\bibitem[{\citenamefont{Avetissian et~al.}(2013)\citenamefont{Avetissian,
  Mkrtchian, Batrakov, Maksimenko, and Hoffmann}}]{Avetissian13}
\bibinfo{author}{\bibfnamefont{H.~K.} \bibnamefont{Avetissian}},
  \bibinfo{author}{\bibfnamefont{G.~F.} \bibnamefont{Mkrtchian}},
  \bibinfo{author}{\bibfnamefont{K.~G.} \bibnamefont{Batrakov}},
  \bibinfo{author}{\bibfnamefont{S.~A.} \bibnamefont{Maksimenko}},
  \bibnamefont{and} \bibinfo{author}{\bibfnamefont{A.}~\bibnamefont{Hoffmann}},
  \bibinfo{journal}{Phys. Rev. B} \textbf{\bibinfo{volume}{88}},
  \bibinfo{pages}{165411} (\bibinfo{year}{2013}).

\bibitem[{\citenamefont{Mikhailov}(2013)}]{Mikhailov13c}
\bibinfo{author}{\bibfnamefont{S.~A.} \bibnamefont{Mikhailov}}, in
  \emph{\bibinfo{booktitle}{Carbon nanotubes and graphene for photonic
  applications}}, edited by
  \bibinfo{editor}{\bibfnamefont{S.}~\bibnamefont{Yamashita}},
  \bibinfo{editor}{\bibfnamefont{Y.}~\bibnamefont{Saito}}, \bibnamefont{and}
  \bibinfo{editor}{\bibfnamefont{J.~H.} \bibnamefont{Choi}}
  (\bibinfo{publisher}{Woodhead Publishing Limited}, \bibinfo{address}{Oxford,
  Cambridge, Philadelphia, New Delhi}, \bibinfo{year}{2013}),
  chap.~\bibinfo{chapter}{7}, pp. \bibinfo{pages}{171--219}.

\bibitem[{\citenamefont{Cheng et~al.}(2014{\natexlab{a}})\citenamefont{Cheng,
  Vermeulen, and Sipe}}]{Cheng14a}
\bibinfo{author}{\bibfnamefont{J.~L.} \bibnamefont{Cheng}},
  \bibinfo{author}{\bibfnamefont{N.}~\bibnamefont{Vermeulen}},
  \bibnamefont{and} \bibinfo{author}{\bibfnamefont{J.~E.} \bibnamefont{Sipe}},
  \bibinfo{journal}{New J. Phys.} \textbf{\bibinfo{volume}{16}},
  \bibinfo{pages}{053014} (\bibinfo{year}{2014}{\natexlab{a}}).

\bibitem[{\citenamefont{Cheng et~al.}(2014{\natexlab{b}})\citenamefont{Cheng,
  Vermeulen, and Sipe}}]{Cheng14b}
\bibinfo{author}{\bibfnamefont{J.~L.} \bibnamefont{Cheng}},
  \bibinfo{author}{\bibfnamefont{N.}~\bibnamefont{Vermeulen}},
  \bibnamefont{and} \bibinfo{author}{\bibfnamefont{J.~E.} \bibnamefont{Sipe}},
  \bibinfo{journal}{Optics Express} \textbf{\bibinfo{volume}{22}},
  \bibinfo{pages}{15868} (\bibinfo{year}{2014}{\natexlab{b}}).

\bibitem[{\citenamefont{Yao et~al.}(2014)\citenamefont{Yao, Tokman, and
  Belyanin}}]{Yao14}
\bibinfo{author}{\bibfnamefont{X.}~\bibnamefont{Yao}},
  \bibinfo{author}{\bibfnamefont{M.}~\bibnamefont{Tokman}}, \bibnamefont{and}
  \bibinfo{author}{\bibfnamefont{A.}~\bibnamefont{Belyanin}},
  \bibinfo{journal}{Phys. Rev. Lett.} \textbf{\bibinfo{volume}{112}},
  \bibinfo{pages}{055501} (\bibinfo{year}{2014}).

\bibitem[{\citenamefont{Smirnova et~al.}(2014)\citenamefont{Smirnova,
  Shadrivov, Miroshnichenko, Smirnov, and Kivshar}}]{Smirnova14}
\bibinfo{author}{\bibfnamefont{D.~A.} \bibnamefont{Smirnova}},
  \bibinfo{author}{\bibfnamefont{I.~V.} \bibnamefont{Shadrivov}},
  \bibinfo{author}{\bibfnamefont{A.~E.} \bibnamefont{Miroshnichenko}},
  \bibinfo{author}{\bibfnamefont{A.~I.} \bibnamefont{Smirnov}},
  \bibnamefont{and} \bibinfo{author}{\bibfnamefont{Y.~S.}
  \bibnamefont{Kivshar}}, \bibinfo{journal}{Phys. Rev. B}
  \textbf{\bibinfo{volume}{90}}, \bibinfo{pages}{035412}
  (\bibinfo{year}{2014}).

\bibitem[{\citenamefont{Peres et~al.}(2014)\citenamefont{Peres, Bludov, Santos,
  Jauho, and Vasilevskiy}}]{Peres14}
\bibinfo{author}{\bibfnamefont{N.~M.~R.} \bibnamefont{Peres}},
  \bibinfo{author}{\bibfnamefont{Y.~V.} \bibnamefont{Bludov}},
  \bibinfo{author}{\bibfnamefont{J.~E.} \bibnamefont{Santos}},
  \bibinfo{author}{\bibfnamefont{A.-P.} \bibnamefont{Jauho}}, \bibnamefont{and}
  \bibinfo{author}{\bibfnamefont{M.~I.} \bibnamefont{Vasilevskiy}},
  \bibinfo{journal}{Phys. Rev. B} \textbf{\bibinfo{volume}{90}},
  \bibinfo{pages}{125425} (\bibinfo{year}{2014}).

\bibitem[{\citenamefont{Cox and {de Abajo}}(2014)}]{CoxAbajo14}
\bibinfo{author}{\bibfnamefont{J.~D.} \bibnamefont{Cox}} \bibnamefont{and}
  \bibinfo{author}{\bibfnamefont{F.~J.~G.} \bibnamefont{{de Abajo}}},
  \bibinfo{journal}{Nat. Commun.} \textbf{\bibinfo{volume}{5}},
  \bibinfo{pages}{5725} (\bibinfo{year}{2014}).

\bibitem[{\citenamefont{Cox and {de Abajo}}(2015)}]{CoxAbajo15}
\bibinfo{author}{\bibfnamefont{J.~D.} \bibnamefont{Cox}} \bibnamefont{and}
  \bibinfo{author}{\bibfnamefont{F.~J.~G.} \bibnamefont{{de Abajo}}},
  \bibinfo{journal}{ACS Photonics} \textbf{\bibinfo{volume}{2}},
  \bibinfo{pages}{306} (\bibinfo{year}{2015}).

\bibitem[{\citenamefont{Savostianova and Mikhailov}(2015)}]{Savostianova15}
\bibinfo{author}{\bibfnamefont{N.~A.} \bibnamefont{Savostianova}}
  \bibnamefont{and} \bibinfo{author}{\bibfnamefont{S.~A.}
  \bibnamefont{Mikhailov}}, \bibinfo{journal}{Appl. Phys. Lett.}
  \textbf{\bibinfo{volume}{107}}, \bibinfo{pages}{181104}
  (\bibinfo{year}{2015}).

\bibitem[{\citenamefont{Cheng et~al.}(2015)\citenamefont{Cheng, Vermeulen, and
  Sipe}}]{Cheng15}
\bibinfo{author}{\bibfnamefont{J.~L.} \bibnamefont{Cheng}},
  \bibinfo{author}{\bibfnamefont{N.}~\bibnamefont{Vermeulen}},
  \bibnamefont{and} \bibinfo{author}{\bibfnamefont{J.~E.} \bibnamefont{Sipe}},
  \bibinfo{journal}{Phys. Rev. B} \textbf{\bibinfo{volume}{91}},
  \bibinfo{pages}{235320} (\bibinfo{year}{2015}).

\bibitem[{\citenamefont{Cheng et~al.}(2016)\citenamefont{Cheng, Vermeulen, and
  Sipe}}]{Cheng16}
\bibinfo{author}{\bibfnamefont{J.~L.} \bibnamefont{Cheng}},
  \bibinfo{author}{\bibfnamefont{N.}~\bibnamefont{Vermeulen}},
  \bibnamefont{and} \bibinfo{author}{\bibfnamefont{J.~E.} \bibnamefont{Sipe}},
  \bibinfo{journal}{Phys. Rev. B} \textbf{\bibinfo{volume}{93}},
  \bibinfo{pages}{039904(E)} (\bibinfo{year}{2016}).

\bibitem[{\citenamefont{Mikhailov}(2016{\natexlab{a}})}]{Mikhailov16a}
\bibinfo{author}{\bibfnamefont{S.~A.} \bibnamefont{Mikhailov}},
  \bibinfo{journal}{Phys. Rev. B} \textbf{\bibinfo{volume}{93}},
  \bibinfo{pages}{085403} (\bibinfo{year}{2016}{\natexlab{a}}).

\bibitem[{\citenamefont{Mikhailov et~al.}(2016)\citenamefont{Mikhailov,
  Savostianova, and Moskalenko}}]{Mikhailov16b}
\bibinfo{author}{\bibfnamefont{S.~A.} \bibnamefont{Mikhailov}},
  \bibinfo{author}{\bibfnamefont{N.~A.} \bibnamefont{Savostianova}},
  \bibnamefont{and} \bibinfo{author}{\bibfnamefont{A.~S.}
  \bibnamefont{Moskalenko}}, \bibinfo{journal}{Phys. Rev. B}
  \textbf{\bibinfo{volume}{94}}, \bibinfo{pages}{035439}
  (\bibinfo{year}{2016}).

\bibitem[{\citenamefont{Rostami and Polini}(2016)}]{Rostami16}
\bibinfo{author}{\bibfnamefont{H.}~\bibnamefont{Rostami}} \bibnamefont{and}
  \bibinfo{author}{\bibfnamefont{M.}~\bibnamefont{Polini}},
  \bibinfo{journal}{Phys. Rev. B} \textbf{\bibinfo{volume}{93}},
  \bibinfo{pages}{161411(R)} (\bibinfo{year}{2016}).

\bibitem[{\citenamefont{Cox et~al.}(2016)\citenamefont{Cox, Silviero, and {de
  Abajo}}}]{CoxAbajo16}
\bibinfo{author}{\bibfnamefont{J.~D.} \bibnamefont{Cox}},
  \bibinfo{author}{\bibfnamefont{I.}~\bibnamefont{Silviero}}, \bibnamefont{and}
  \bibinfo{author}{\bibfnamefont{F.~J.~G.} \bibnamefont{{de Abajo}}},
  \bibinfo{journal}{ACS NANO} \textbf{\bibinfo{volume}{10}},
  \bibinfo{pages}{1995} (\bibinfo{year}{2016}).

\bibitem[{\citenamefont{Marini et~al.}(2016)\citenamefont{Marini, Cox, and {de
  Abajo}}}]{MariniAbajo16}
\bibinfo{author}{\bibfnamefont{A.}~\bibnamefont{Marini}},
  \bibinfo{author}{\bibfnamefont{J.~D.} \bibnamefont{Cox}}, \bibnamefont{and}
  \bibinfo{author}{\bibfnamefont{F.~J.~G.} \bibnamefont{{de Abajo}}},
  \bibinfo{journal}{arXiv:1605.06499}  (\bibinfo{year}{2016}).

\bibitem[{\citenamefont{Sharif et~al.}(2016)\citenamefont{Sharif, Ara, Ghafary,
  Salmani, and Mohajer}}]{Sharif16}
\bibinfo{author}{\bibfnamefont{M.~A.} \bibnamefont{Sharif}},
  \bibinfo{author}{\bibfnamefont{M.~H.~M.} \bibnamefont{Ara}},
  \bibinfo{author}{\bibfnamefont{B.}~\bibnamefont{Ghafary}},
  \bibinfo{author}{\bibfnamefont{S.}~\bibnamefont{Salmani}}, \bibnamefont{and}
  \bibinfo{author}{\bibfnamefont{S.}~\bibnamefont{Mohajer}},
  \bibinfo{journal}{Opt. Mater.} \textbf{\bibinfo{volume}{53}},
  \bibinfo{pages}{80} (\bibinfo{year}{2016}).

\bibitem[{\citenamefont{Mikhailov}(2016{\natexlab{b}})}]{Mikhailov16c}
\bibinfo{author}{\bibfnamefont{S.~A.} \bibnamefont{Mikhailov}}
  (\bibinfo{year}{2016}{\natexlab{b}}), \bibinfo{note}{arXiv:1608.00877}.

\bibitem[{\citenamefont{Glazov and Ganichev}(2014)}]{Glazov14}
\bibinfo{author}{\bibfnamefont{M.~M.} \bibnamefont{Glazov}} \bibnamefont{and}
  \bibinfo{author}{\bibfnamefont{S.}~\bibnamefont{Ganichev}},
  \bibinfo{journal}{Phys. Rep.} \textbf{\bibinfo{volume}{535}},
  \bibinfo{pages}{101} (\bibinfo{year}{2014}).

\bibitem[{\citenamefont{Hartmann et~al.}(2014)\citenamefont{Hartmann, Kono, and
  Portnoi}}]{Hartmann14}
\bibinfo{author}{\bibfnamefont{R.~R.} \bibnamefont{Hartmann}},
  \bibinfo{author}{\bibfnamefont{J.}~\bibnamefont{Kono}}, \bibnamefont{and}
  \bibinfo{author}{\bibfnamefont{M.~E.} \bibnamefont{Portnoi}},
  \bibinfo{journal}{Nanotechnology} \textbf{\bibinfo{volume}{25}},
  \bibinfo{pages}{322001} (\bibinfo{year}{2014}).

\bibitem[{\citenamefont{Mikhailov and Ziegler}(2007)}]{Mikhailov07d}
\bibinfo{author}{\bibfnamefont{S.~A.} \bibnamefont{Mikhailov}}
  \bibnamefont{and} \bibinfo{author}{\bibfnamefont{K.}~\bibnamefont{Ziegler}},
  \bibinfo{journal}{Phys. Rev. Lett.} \textbf{\bibinfo{volume}{99}},
  \bibinfo{pages}{016803} (\bibinfo{year}{2007}).

\bibitem[{\citenamefont{Schubert et~al.}(2000)\citenamefont{Schubert, Tiwald,
  and Herzinger}}]{Schubert00}
\bibinfo{author}{\bibfnamefont{M.}~\bibnamefont{Schubert}},
  \bibinfo{author}{\bibfnamefont{T.~E.} \bibnamefont{Tiwald}},
  \bibnamefont{and} \bibinfo{author}{\bibfnamefont{C.~M.}
  \bibnamefont{Herzinger}}, \bibinfo{journal}{Phys. Rev. B}
  \textbf{\bibinfo{volume}{61}}, \bibinfo{pages}{8187} (\bibinfo{year}{2000}).

\bibitem[{\citenamefont{Johnson and Christy}(1972)}]{Johnson72}
\bibinfo{author}{\bibfnamefont{P.~B.} \bibnamefont{Johnson}} \bibnamefont{and}
  \bibinfo{author}{\bibfnamefont{R.~W.} \bibnamefont{Christy}},
  \bibinfo{journal}{Phys. Rev. B} \textbf{\bibinfo{volume}{6}},
  \bibinfo{pages}{4370} (\bibinfo{year}{1972}).

\bibitem[{\citenamefont{Olmon et~al.}(2012)\citenamefont{Olmon, Slovick,
  Johnson, Shelton, Oh, Boreman, and Raschke}}]{Olmon12}
\bibinfo{author}{\bibfnamefont{R.~L.} \bibnamefont{Olmon}},
  \bibinfo{author}{\bibfnamefont{B.}~\bibnamefont{Slovick}},
  \bibinfo{author}{\bibfnamefont{T.~W.} \bibnamefont{Johnson}},
  \bibinfo{author}{\bibfnamefont{D.}~\bibnamefont{Shelton}},
  \bibinfo{author}{\bibfnamefont{S.-H.} \bibnamefont{Oh}},
  \bibinfo{author}{\bibfnamefont{G.~D.} \bibnamefont{Boreman}},
  \bibnamefont{and} \bibinfo{author}{\bibfnamefont{M.~B.}
  \bibnamefont{Raschke}}, \bibinfo{journal}{Phys. Rev. B}
  \textbf{\bibinfo{volume}{86}}, \bibinfo{pages}{235147}
  (\bibinfo{year}{2012}).

\end{thebibliography}

\end{document}